\documentclass[11pt,a4paper]{article}

\usepackage{jheppub}
\usepackage{epsfig,latexsym,amsfonts,amsmath,amsthm,amssymb,amsbsy,multirow,slashed,color,mathrsfs,wasysym,textcomp,subfigure,wrapfig,comment}
\usepackage{hyperref}
\hypersetup{colorlinks,
 linkcolor = blue,
 anchorcolor = red,
 citecolor = blue,
 filecolor = red,
 urlcolor = blue,
 linktocpage}
\usepackage{color,datetime,graphicx}
\usepackage{float}
\usepackage{subfigure} 

\newcommand{\be}{\begin{equation}}
\newcommand{\ee}{\end{equation}}
\newcommand{\beqa}{\begin{eqnarray}}
\newcommand{\eeqa}{\end{eqnarray}}

\newcommand{\pink}[1]{\textcolor{\pink}{#1}}

\title{Localised Anti-Branes in Flux Backgrounds}
\preprint{}

\author{Gavin S. Hartnett}
\affiliation{Department of Physics, UCSB, Santa Barbara, CA 93106}
\emailAdd{hartnett@physics.ucsb.edu}

\abstract{
Solutions corresponding to finite temperature (anti)-D3 and M2 branes localised in flux backgrounds are constructed in a linear approximation. The flux backgrounds considered are toy models for the IR of the Klebanov-Strassler solution and its M-theory analogue, the Cveti\v{c}-Gibbons-L\"{u}-Pope solution. Smooth solutions exist for either sign charge, in stark contrast with the previously considered case of smeared black branes. That the singularities of the anti-branes in the zero temperature extremal limit can be shielded behind a finite temperature horizon indicates that the singularities are physical and resolvable by string theory. As the charge of the branes grows large and negative, the flux at the horizon increases without bound and diverges in the extremal limit, which suggests a resolution via brane polarisation \`{a} la Polchinski-Strassler. It therefore appears that the anti-brane singularities do not indicate a problem with the SUSY-breaking metastable states corresponding to expanded anti-brane configurations in these backgrounds, nor with the use of these states in constructing the de Sitter landscape.
}

\begin{document}
\maketitle

\section{Introduction}
There has been much recent work on the study of anti-branes in flux backgrounds.\footnote{For a recent review, including a comprehensive list of references, see \cite{DanielssonXYZ}.} This interest is historically motivated by the work of Kachru, Pearson, and Verlinde (KPV) \cite{Kachru:2002gs} who found that a small number of anti-D3 branes placed at the tip of the Klebanov-Strassler (KS) solution \cite{Klebanov:2000hb} would expand into a metastable NS5 brane. The NS5 brane would then eventually decay via tunnelling, although its lifetime could be tuned to be parametrically long. This mechanism has led to a number of interesting applications, including the construction of de Sitter (dS) compactifications of string theory in the KKLT scenario \cite{Kachru:2003aw}, holographic duals of dynamical supersymmetry breaking \cite{Kachru:2003aw, Argurio:2007qk, Argurio:2006ny}, and non-extremal black hole microstate geometries \cite{Bena:2011fc, Bena:2012zi}.

Given these varied and important applications, it is of obvious interest to construct the full backreacted NS5 brane solution and to study its properties. Unfortunately, such a solution would depend non-trivially on two coordinates, leading to a complicated non-linear PDE problem that is quite intractable. As a result, the backreaction of smeared anti-D3 branes added to the tip of the KS solution was studied instead, and unexpected singularities in the 3-form field strengths were found at the location of the anti-branes \cite{McGuirk:2009xx, Bena:2009xk, Bena:2011hz, Massai:2012jn, Bena:2012bk}. This prompted many additional investigations, the results of which all point to the fact that singularities generically arise when anti-branes are added to flux backgrounds in both string theory and M-theory \cite{Blaback:2011nz, Blaback:2011pn, Blaback:2014tfa}.\footnote{Here and throughout this article, by anti-brane we mean that the brane is not mutually BPS with respect to the flux background.} Importantly, these studies also found that the singularities are not artefacts of either linearisation in the anti-brane charge nor smearing. 

The physical interpretation of these singularities has been the source of much debate. String theory is known to resolve many singularities in gravity. Perhaps one of the most well-known mechanisms for resolving singularities is through brane polarisation, also known as the Myers effect \cite{Myers:1999ps}, which was shown to resolve the singularity associated with a mass deformation of $\mathcal{N}=1^*$ $SU(N)$ supersymmetric Yang-Mills by Polchinski and Strassler (PS) \cite{Polchinski:2000uf}. Another famous example concerns the KS solution itself, which can be interpreted as the resolution of the singularity present in the Klebanov-Tseytlin \cite{Klebanov:2000nc} solution. Not all singularities are resolved by string theory however, and in fact singular solutions play an important role in constraining the theory by, for example, ruling out negative mass states \cite{Horowitz:1995ta}. 

The singularities associated with anti-branes in flux backgrounds throws the fate of the KPV metastable state and its many applications into doubt. Thus far, all attempts to resolve the singularities via polarisation have either failed \cite{Bena:2012tx, Bena:2012vz}, or partially succeeded \cite{Bena:2014bxa, Bena:2014jaa}, but have also indicated the presence of repulsive instabilities which might destabilize the metastable states.\footnote{An example where polarisation does resolves the singularity without any repulsive instability is \cite{Junghans:2014wda} for the case of $AdS_7$ flux vacua. The polarisation relied on the curvature of the worldvolume $AdS_7$, and the flux backgrounds we consider here will have no curvature on the worldvolume.} As the backreaction of localised anti-branes in flux backgrounds has been too difficult to tackle directly, these authors cleverly extracted the polarisation potentials from the backreacted smeared solutions.

A widely held view, dubbed the Gubser criterion \cite{Gubser:2000nd}, is that if a singularity can be shielded behind a finite temperature horizon, then it is physical and resolvable by some (perhaps unknown) mechanism in string theory. According to this criterion, there is then an important relationship between the space of black brane solutions and anti-brane singularities. For a horizon to shield an anti-brane singularity the solution should asymptote to the flux background of interest, for example KS, and the horizon should carry negative charge. Thus far all attempts (including both analytic arguments and the numerical construction of solutions) to hide \textit{smeared} anti-brane singularities behind horizons have failed, and only black branes with positive charge have been found \cite{Bena:2012ek, Bena:2013hr}. Additionally, a no-go theorem has been recently formulated which excludes non-singular solutions corresponding to anti-branes in flux backgrounds at both zero and finite temperature, regardless of smearing \cite{Blaback:2014tfa}. 

If the interpretation of the smeared anti-brane singularities is that they are unphysical and unable to be resolved as these many results suggest, then there are dramatic consequences for the de Sitter landscape, metastable non-SUSY field theory states, and the construction of non-extremal microstate geometries. It also raises a vexing contradiction with the original KPV calculation \cite{Kachru:2002gs}--namely, where did this calculation go wrong? One interesting suggestion is that the probe approximation severely underestimated the amount of flux clumping caused by charge screening, and that the branes annihilate against the flux through classical time evolution \cite{Blaback:2011pn, Blaback:2012nf, Danielsson:2014yga}.

In this work, we will provide evidence that the metastable states do in fact exist, and show that the singularity of anti-branes in flux backgrounds can be resolved. We will accomplish this by constructing, in a linear approximation, localised black branes of either sign charge in flux backgrounds that approximate the IR of KS in Type IIB string theory, and the analogous solution in M-theory, the Cveti\v{c}-Gibbons-L\"{u}-Pope (CGLP) solution \cite{Cvetic:2000db}. At first glance this result appears to directly contradict the no-go theorem of \cite{Blaback:2014tfa}, however it has a loop-hole which our constructions make use of. The existence of these solutions demonstrates that the anti-brane singularities can be shielded behind a smooth horizon. According to the Gubser criterion then, \textit{the anti-brane singularities are physical and are resolved by string theory.} It therefore appears that, in contrast with the title of \cite{Bena:2012ek}, horizons can and do in fact save the landscape, once the smearing approximation is dropped!\footnote{It should be emphasized that the anti-brane singularities themselves are not artefacts of smearing.}

Insight into the nature of the resolution can be gained by considering the extremal limit. As the charge is made large and negative, the flux at the horizon grows without bound. In the extremal anti-brane limit, the flux is found to diverge in a way known to be resolved in certain cases by brane polarisation in a non-supersymmetric version of Polchinski-Strassler. Although we cannot directly evaluate this possibility without going to higher order in our approximation scheme, we will argue that indeed polarisation resolves the singularity. As mentioned above, this resolution mechanism was previously argued to not apply \cite{Bena:2012tx, Bena:2012vz, Bena:2014bxa, Bena:2014jaa}, and we will therefore attempt to explain the discrepancy as a consequence of smearing. If, however, polarisation does not resolve the singularity, then either there is some new resolution mechanism yet to be discovered, or else the Gubser criterion fails and being able to shield a singularity behind a horizon is not a sufficient criterion for accepting it as physical.

The approach taken here, where the number of anti-branes is taken to be large enough that the gravity description is valid, is complementary to the recent analysis of \cite{Michel:2014lva}, who argued that for at least a single anti-brane, the physics could be well understood from the effective field theory of the brane worldvolume theory. It therefore seems that for both a single anti-brane and many, KKLT \cite{Kachru:2003aw} remains a valid mechanism for constructing metastable dS vacua in string theory.

This paper is organized as follows. In Sec.~\ref{sec:IIB} we construct solutions corresponding to D3 branes at both finite and zero temperature localised in a toy flux background which approximates the IR of the KS solution. The analogous problem of anti-M2 branes in a toy background which approximates the CGLP solution of M-theory is quite similar, and all the main conclusions of the Type IIB case carry over. Accordingly, we relegate its treatment to Appendix \ref{sec:M-theory}. We close with concluding remarks in  Sec.~\ref{sec:conclusion}.

\section{D3 Branes in Type IIB Flux Backgrounds
\label{sec:IIB}}
One of the major obstacles in the problem of anti-D3 branes in KS has been the complicated geometry of the deformed conifold which has discouraged attempts to study inhomogeneous or localised solutions. Therefore, in this section we will study localised D3 and anti-D3 branes in a toy flux background which approximates the IR of the KS solution. As we will argue below, this serves as an excellent approximation for the solution corresponding to branes at the tip of the KS solution. We first introduce the background and argue that it can be used to study the problem of anti-branes in KS, and then review smeared solutions before finally constructing localised solutions.

\subsection{A Toy Model of the Klebanov-Strassler Flux Background}
The Klebanov-Strassler solution \cite{Klebanov:2000hb} is the supersymmetric Type IIB supergravity solution corresponding to fractional D3 branes at the tip of the deformed conifold. Recall that the conifold is a six-dimensional Ricci-flat space which asymptotically takes the form of the cone $ds_6^2 \sim dr^2 + r^2 ds^2_{\mathbb{T}^{1,1}}$\footnote{Here $\mathbb{T}^{1,1}$ is the coset space $SU(2)\times SU(2)/U(1)$.}, and smoothly caps off at the tip, where the geometry consists of a finite-sized $S^3$ and a shrinking $S^2$. A simple exact solution of Type IIB supergravity which serves as a toy model relevant for approximating the IR of KS is
\begin{align} 
\label{eq:IIBflux}
ds^2 &= H^{-1/2} \left(-dt^2 + dx_1^2 + dx_2^2 + dx_3^2 \right) + H^{1/2}\left( dr_1^2 + r_1^2 d\Omega_2^2 + \sum_{i=1}^3 dy_i^2 \right), \\
F_5 &= (1+\star)\mathcal{F}_5, \qquad \mathcal{F}_5 = \varepsilon_F dH^{-1} \wedge dt \wedge dx^1 \wedge dx^2 \wedge dx^3, \nonumber \\
G_3 &= F_3 - i H_3 = m \left( dy^1 \wedge dy^2 \wedge dy^3 - i \varepsilon_F  r_1^2 dr_1 \wedge d\Omega_2 \right), \nonumber
\end{align}
with all other fields zero. The warp factor is
\begin{equation} 
H = 1 + \frac{a_0}{r_1} - \frac{m^2}{6} r_1^2. 
\end{equation}
The complex 3-form $G_3$ is (anti)-imaginary self-dual, that is $\star_6 G_3 = i \varepsilon_F  G_3$, where $\star_6$ is the Hodge dual on the 6-dimensional transverse space. A solution exists for either sign of the flux, $\varepsilon_F = \pm 1$, and the constant $a_0$ is proportional to the number of (anti)-D3 branes smeared over the $y$ directions, which for the purpose of approximating KS, we set to zero, $a_0 = 0$. The constant $m$ controls the amount of flux and is analogous to the $M$ constant in the KS solution. This solution is related via T-duality to the ``massive-D6'' solution of massive IIA supergravity found in \cite{Janssen:1999sa}, and was used as a toy model for KS in the context of adding smeared anti-branes to the solution by  \cite{Bena:2012tx, Bena:2013hr}. 

That this solution approximates the IR of KS can be seen as follows. The geometry near the ``tip'', $r_1 \rightarrow 0$, consists of a shrinking $S^2$ times $\mathbb{R}^3$, which is very similar to the shrinking $S^2$ times $S^3$ geometry of the near-tip KS solution. In both solutions $F_3$ near the tip is simply the volume form on the non-shrinking space, either $S^3$ or $\mathbb{R}^3$. Also, for each solution $H_3$ contains a term proportional to the volume form on the shrinking space (in the toy flux background, this is the only term present). Therefore, the IR of both solutions is seen to be quite similar--the essential difference is that one solution contains an $S^3$ at the tip and the other an $\mathbb{R}^3$.

In this paper we will be primarily interested in approximating the solution corresponding to anti-D3 branes localised at both the tip of the deformed conifold and at a point on the $S^3$. Since the branes sit at a single point, the curvature of the $S^3$ is negligible, and therefore the near-brane solution will be very similar to the analogous solution in the toy flux background \eqref{eq:IIBflux}. If the branes are blackened by turning on a small temperature, then they will no longer be localised at a single point, but for sufficiently small temperature their Schwarzschild radius will be small compared to the curvature of the $S^3$ and the toy flux background will remain an excellent approximation for KS in the vicinity of the branes. For the sake of constructing these solutions, it will be convenient to introduce spherical coordinates for the $y^i$,
\begin{equation}
\sum_{i=1}^3 dy_i^2 = dr_2^2 + r_2^2 d\widetilde{\Omega}_2^2, \qquad dy^1 \wedge dy^2 \wedge dy^3 = r_2^2 dr_2 \wedge d\widetilde{\Omega}_2,
\end{equation}
where $d\widetilde{\Omega}_2^2$, $d\widetilde{\Omega}_2$ are the line element and volume form for a second, distinct unit $S^2$.

The toy flux backgroud \eqref{eq:IIBflux} has a disturbing property. For large $r_1$, the warp factor $H$ is large and negative, indicating that at some point it passes through zero. For the case of interest, $a_0 = 0$, this occurs for $r_1 = \sqrt{6}/m$. We will argue that this naked singularity and strange asymptotics are actually  irrelevant for the present purpose of studying localised anti-branes in flux backgrounds. Firstly, the singularity occurs at a radius that can be taken much larger than the radius of the localised solution, and for small radii the toy flux background approximates the KS solution. Secondly, it is shown in Appendix~\ref{sec:naked} that although the supergravity description is singular, strings are in fact well behaved and perfectly non-singular at the point $H=0$ \footnote{This point was understood as a result of joint work with G. Horowitz and A. Puhm.}. 

\subsection{Analogue of the KPV Calculation
\label{subsec:IIBpolarisation}}
In this section we briefly review the KPV \cite{Kachru:2002gs} calculation, in which a metastable NS5 brane is found to exist in the KS solution. We then repeat the analysis for the toy flux background introduced above, \eqref{eq:IIBflux}. Although there are important differences between the two cases, the main conclusion remains the same, namely that the probe calculation indicates that a static, expanded brane configuration should exist, strengthening the argument of the previous section that localised branes in the IR of KS are be well modelled by the analogous solution in the toy flux background.

\begin{figure}[H]
\begin{center}
\includegraphics[width=0.5\textwidth]{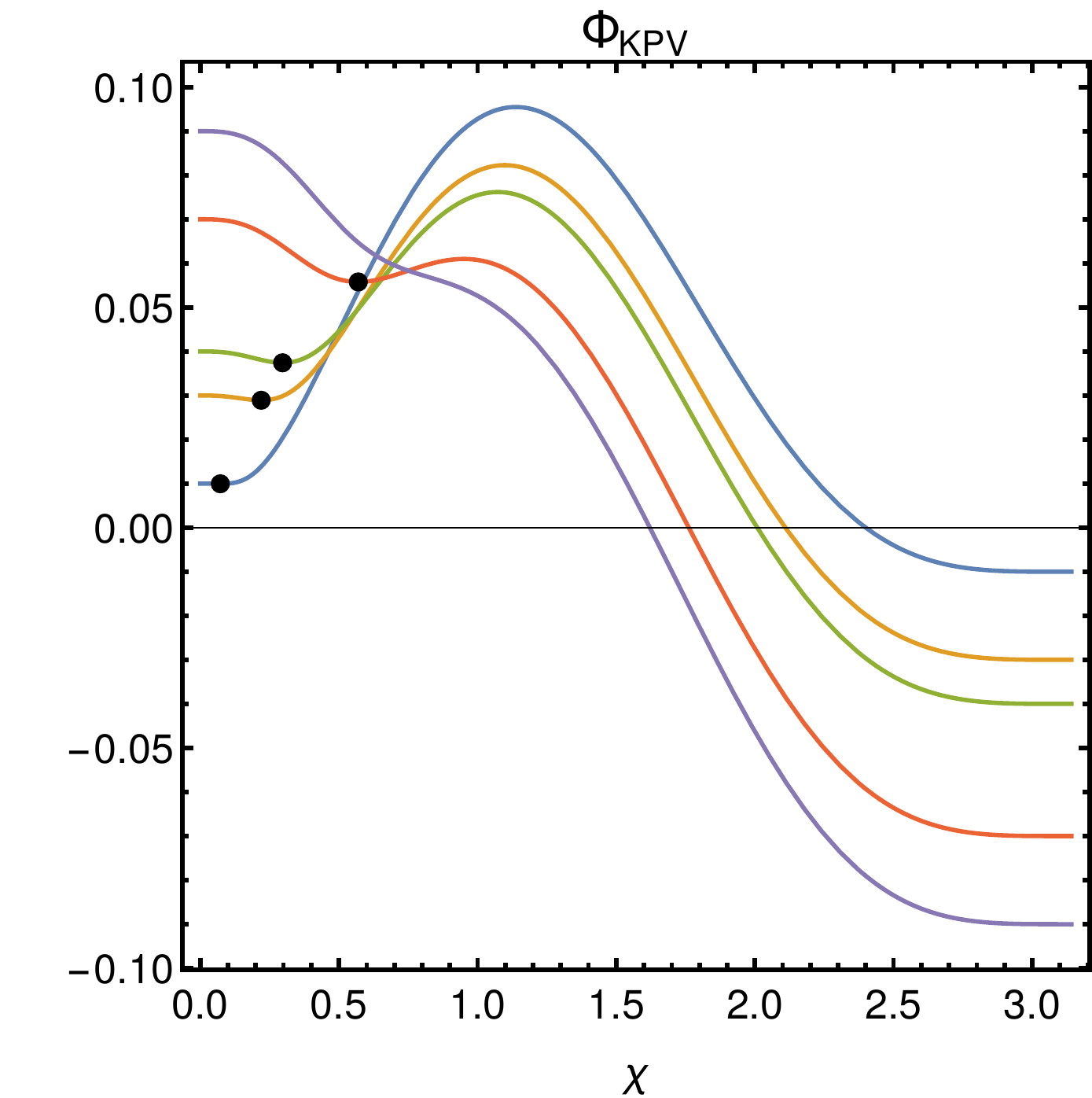}
\end{center}
\caption{$p/M = 0.01, 0.03, 0.04, 0.07, 0.09$, from bottom to top. Metastable minima exist for $p/M \lesssim 0.08$, and are indicated by black dots. 
\label{fig:KPVplot}}
\end{figure}

In KPV, the potential for a probe NS5 brane carrying $p$ units of anti-D3 charge was computed. The brane was embedded at the tip of the KS flux background as follows: it filled out the four-dimensional Minkowski space and wrapped an $S^2$ inside the finitely sized $S^3$ at the tip. Writing the line element on the $S^3$ as $d\chi^2 + \sin^2\chi^2 d\Omega_2^2$, the NS5 brane was localised at fixed $\chi$, and the potential was found in \cite{Kachru:2002gs} to be proportional to
\begin{equation}
\Phi_{\text{KPV}}(\chi) = \frac{1}{\pi}\left[ \sqrt{b_0^4 \sin^4\chi + \left(\pi \frac{p}{M} -\chi + \frac{1}{2}\sin2\chi \right)^2} - \chi + \frac{1}{2}\sin2\chi \right], 
\end{equation}
where $b_0^2 \approx 0.93266$, $p$ is the anti-D3 charge, and $M$ is the KS flux parameter. This potential is plotted in Fig.~\ref{fig:KPVplot} for various values of $p$.

For anti-branes $(p>0)$, the south pole $\chi = \pi$ is always the global minima. For $p/M \lesssim 0.08$ there is a second, metastable minima corresponding to a puffed-up NS5 brane. As $p$ increases, the size of the wrapped $S^2$ increases from zero until a minima with a maximally-sized $S^2$ is attained for $p/M \simeq 0.08$. Brane flux-annihilation occurs either via tunnelling $(p/M \lesssim 0.08$) or through classical evolution by rolling down the potential hill $(p/M \gtrsim 0.08)$.

For the present case of anti-branes in the toy flux background the analogous set-up would be that the branes are localised ``at the tip'' ($r_1 = 0$), as well as at the origin of the transverse $\mathbb{R}^3$ ($r_2 = 0$). For small enough radii of the wrapped $S^2$, the difference between an $S^2$ in $S^3$ or $\mathbb{R}^3$ should vanish, and the KPV potential should match the potential for the toy background. Since the KPV potential always has a minima in the limit of $p\rightarrow 0$, and in this limit the radius of the wrapped $S^2$ vanishes, the anti-branes at the tip of the KS solution should be excellently approximated by anti-branes at the tip of the toy flux background for $p$ small enough (but large enough so that the gravity description is valid).

Although the potentials should agree for small radius, they will be quite different in the opposite limit of large radius. Since $\mathbb{R}^3$ is topologically trivial, the true minima corresponding to the south pole of the $S^3$ will not exist in the toy flux background, and in fact there should be no analogue of the KPV brane-flux annihilation mechanism where the wrapped $S^2$ tunnels from the metastable minima to the south pole.

\begin{figure}[]
\begin{center}
\includegraphics[width=0.5\textwidth]{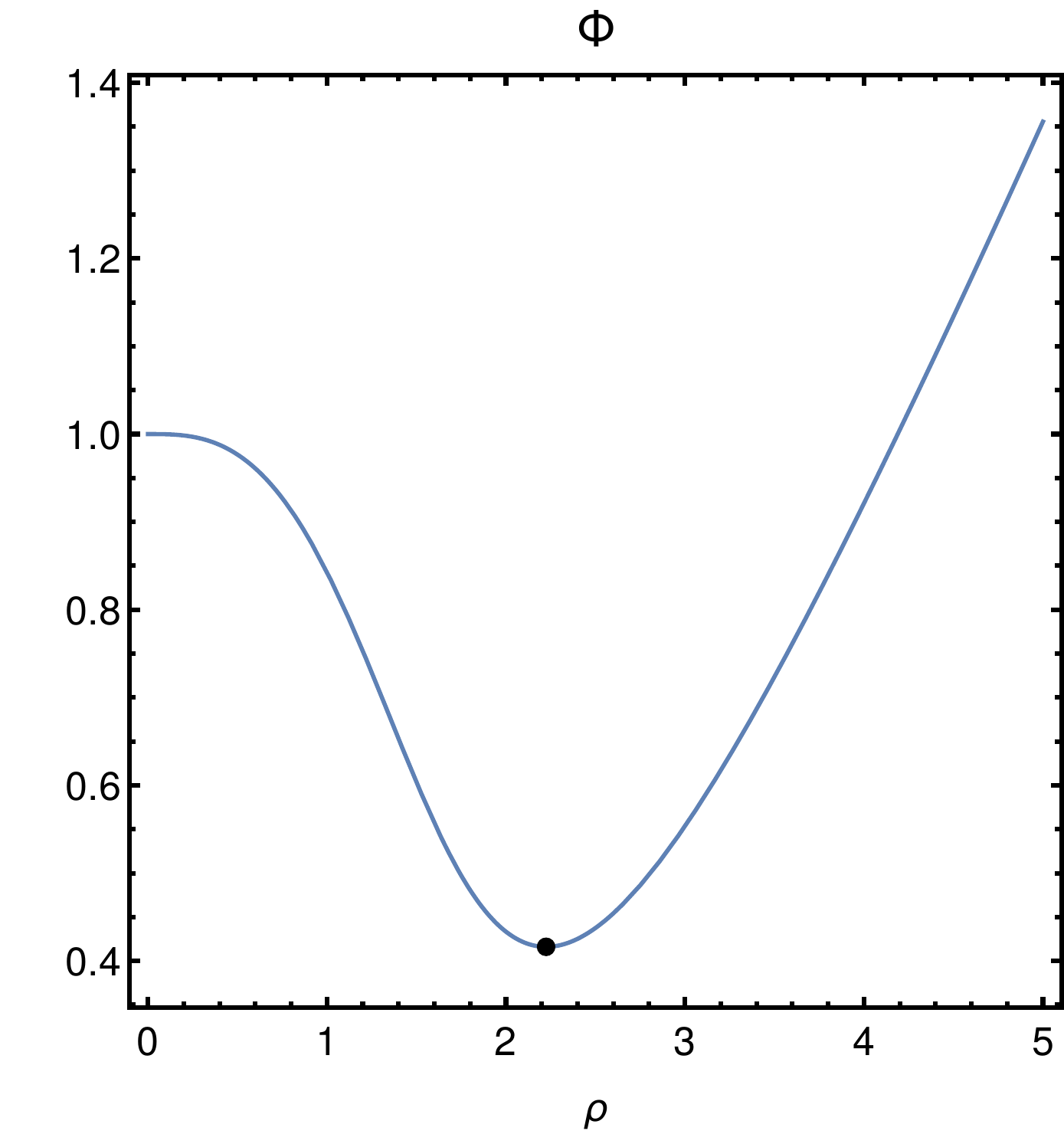}
\end{center}
\caption{The probe potential $\Phi$ for $p/\mu=1$. The global minima is indicated by a black dot. The key differences between the potential for the KS background and the toy flux background \eqref{eq:IIBflux} are that the minima corresponding to an NS5 brane wrapping a finitely sized $S^2$ is 1) always present for the toy flux background, and 2) always the global minima as opposed to simply a metastable minima.\label{fig:toyKPV}}
\end{figure}

Let us now turn to the calculation of the potential. The NS5 brane action is \footnote{Here, and throughout this paper we set $g_s = 1$.}
\begin{equation}
S =  \mu_5 \int d^6 \xi \left(-\det G_{||} \det\left(G_{\perp} + 2\pi \mathcal{F}_2 \right) \right)^{1/2} + \mu_5 \int B_6. 
\end{equation}
Here $2\pi \mathcal{F}_2 = 2\pi F_2 -C_2$, with $F_3 = dC_2 = m d \left(r_2^3/3 \right) \wedge d\widetilde{\Omega}_2$ , and $F_2 = (p/2) d\widetilde{\Omega}_2$ represents $p$ units of anti-D3 charge carried by the NS5 brane worldvolume. The NS5 brane couples to $B_6$, where 
\begin{equation}
\star H_3 = H_7 = dB_6 = -\frac{m \varepsilon_F}{H} d \left(\frac{r_2^3}{3} \right) \wedge d^4 x \wedge d\widetilde{\Omega}_2 .
\end{equation}
The potential is then found to be proportional to
\begin{equation}
\Phi = \frac{m^2}{\pi} \left[ \sqrt{r_2^4 + \left(p \pi - \frac{m r_2^3}{3} \right)^2 } - \frac{m  \varepsilon_F}{3} r_2^3 \right], 
\end{equation}
where the value of the warp factor at the tip, $H(r_1 = 0) = 1$, has been used. To compare with the KPV potential, it is useful to convert to the new flux parameter $m =\mu^{-1/2}$ and dimensionless coordinate $r_2 = \mu^{1/2} \rho$. Then for $\varepsilon_F = 1$, the behaviour of the two potentials near the tip is
\begin{align}
& \Phi_{\text{KPV}} \sim \frac{p}{M} - \frac{4}{3\pi} \chi^3 + \frac{b_0^4}{2\pi^2} \frac{M}{p} \chi^4 + \mathcal{O}(\chi^5), \\
& \Phi \sim \frac{p}{\mu} - \frac{2}{3\pi} \rho^3 + \frac{1}{2\pi^2} \frac{\mu}{p} \rho^4 + \mathcal{O}(\rho^{7}). \nonumber 
\end{align}
There are two sources for the different coefficients. Firstly, deviations should be expected at high enough order for the simple reason that the geometries are different. Secondly, we have not taken care to ensure that the tip of the KS solution is parametrized identically to the tip of the toy flux background. The large $\rho$-behaviour of the toy flux background potential strongly differs from the KPV behaviour, and is
\begin{equation}
\Phi \sim \frac{3}{2\pi} \rho - \frac{p}{\mu} + \mathcal{O}(\rho^{-1}) .
\end{equation}
Since the potential is decreasing near $\rho = 0$, but increasing for large $\rho$, there must be a turning point at a real positive value of $\rho$ \textit{for all values of $p$!} In Fig.~\ref{fig:toyKPV} this potential is plotted for  $p/\mu=1$.

As expected, there are important global differences between the two potentials. The finite $\rho$-minima in the toy background always exists and is the global minima, and there is no possibility of brane-flux annihilation through either quantum tunnelling or classical evolution. However, for small $p$ and small $\rho$, the KPV potential is qualitatively very similar to the potential for the toy background once the coordinates $\chi$ and $\rho$ are identified, and a perturbatively stable, static expanded brane configuration exists in both solutions.

\subsection{Smeared Branes
\label{subsec:IIBsmearing}}
Before turning to the main focus of this paper, the construction of localised black branes in the toy flux background \eqref{eq:IIBflux}, we first discuss smeared solutions. In the extremal case, Ref.~\cite{Bena:2012tx} found that the solution T-dual to smeared anti-D3 branes in the toy flux background possessed singular fluxes which were not resolved by brane polarisation. In the finite-temperature case, Ref.~\cite{Bena:2013hr} formulated a no-go theorem excluding regular black hole solutions where the charge on the black hole is negative. These results strongly suggest that the singularity associated with \textit{smeared} anti-branes in this flux background is simply unphysical and unable to be resolved.

In order to highlight the important differences between smeared and localised solutions, here we will approximately construct the solution corresponding to a positively charged smeared brane in the toy background. Near the brane the solution can be approximated as a linear $G_3$ flux perturbation of the thermal, smeared D3 brane solution:
\begin{align}
\label{eq:D3smeared} 
ds^2 &= H^{-1/2} \left(-f dt^2 + dx_1^2 + dx_2^2 + dx_3^2 \right) + H^{1/2} \left( \frac{dr_1^2}{f} + r_1^2 d\Omega_2^2 + dy_1^2 + dy_2^2 + dy_3^2 \right), \nonumber \\
\mathcal{F}_5 &= \varepsilon_B \coth\beta \, dH^{-1} \wedge dt \wedge dx^1 \wedge dx^2 \wedge dx^3, \\
H &= 1 + \sinh^2 \beta \left( \frac{r_+}{r_1} \right), \qquad f = 1 - \frac{r_+}{r_1}. \nonumber
\end{align}
The boost parameter $\beta$ characterises the charge of the brane, and we will take $\beta \ge 0$ and allow $\varepsilon_B = \pm 1$ to characterise the difference between positive and negative charge. The extremal limit is $\beta \rightarrow \infty$ while keeping $a_0 \equiv r_+ \sinh^2\beta$ fixed.

The most general ansatz for the perturbation consistent with the symmetries of the problem is
\begin{equation}
G_3 = m \left( g_y dy^1 \wedge dy^2 \wedge dy^3 - i \, g_{\Omega} r_1^2 dr_1 \wedge d\Omega_2 \right),
\end{equation}
where $g_y$, $g_{\Omega}$ are functions of $r_1$ only. The Bianchi constraint $dG_3 = 0$ fixes $g_y$ to be a constant, which we choose to be $g_y = 1$. The linearised equation of motion ${d \star G_3 + i G_3 \wedge F_5 = 0}$ then becomes
\begin{equation}
\partial_{r_1} \left(H^{-1} f g_{\Omega} - \varepsilon_B \coth\beta  H^{-1} \right) = 0,
\end{equation}
which has the simple solution
\begin{equation}
\label{eq:IIBsmearedpert}
g_{\Omega} = f^{-1} \left( H c_0 + \varepsilon_B \coth\beta \right),
\end{equation}
with $c_0$ a constant of integration. First, consider the extremal case. The solution is 
\begin{equation}
g_{\Omega} = c_0 \left(1 + \frac{a_0}{r_1} \right) + \varepsilon_B.
\end{equation}
Regularity at the brane location requires $c_0 = 0$, in which case the flux has a definite imaginary self-duality (ISD) character: $\star_6 G_3 = i \varepsilon_B G_3$, where $\star_6$ is  the Hodge operator on the transverse 6-dimensional space. Crucially, the flux is constrained to have the same self-duality sign as the brane charge. In fact, this solution just corresponds to adding ISD flux to the smeared brane solution (or AISD flux to the smeared anti-brane solution), i.e. it is simply a linearisation of the toy flux background \eqref{eq:IIBflux} with the warp factor replaced by
\begin{equation}
H = 1 + \frac{a_0}{r} - \frac{m^2 r_1^2}{6}. 
\end{equation}

Suppose we took the opposite perspective and fixed the asymptotic value of the flux, rather than its behaviour near the brane source. Requiring the flux to have the opposite sign, $\star_6 G_3 = - i \varepsilon_B G_3$, fixes $c_0 = - 2 \varepsilon_B$, which causes the flux to diverge at $r_1 = 0$. This perturbation corresponds to the linearisation of the solution corresponding to anti-branes added to the toy flux background. These singularities were shown to not be resolved by polarisation in \cite{Bena:2012tx}.

Next, consider the finite temperature case. Requiring regularity at the horizon fixes $c_0$ and the solution is simply
\begin{equation}
g_{\Omega}  = \varepsilon_B \tanh\beta.
\end{equation}
This perturbation corresponds to the linearisation of the positively charged smeared black brane in a flux background (which has not been constructed yet). The sign of the flux is constrained to be of the same sign as the brane charge, in agreement with the no-go theorems ruling out negative charged branes \cite{Bena:2013hr}. In fact, the later and more comprehensive no-go theorem of \cite{Blaback:2014tfa} correctly predicts the divergence associated with the negatively charged brane. If the constant $c_0$ had been taken to be any other value, including values that would yield negatively charged black branes, then the 3-form fluxes would diverge as $|H_3|^2 \sim f^{-1}$. This point will be important later.

Having reviewed the problem of smeared anti-branes in the toy background \eqref{eq:IIBflux}, we now turn to the central focus of this paper, the construction of localised black branes. 

\subsection{Localised Branes
\label{subsec:IIBlocalised}}
Here we will construct the solution corresponding to localised D3 branes at the tip of the toy flux background \eqref{eq:IIBflux} to linear order in the 3-form flux. Since the backreaction of the flux alters the asymptotics, this approximation is only valid in the vicinity of the branes. The approach we take is very similar to the one employed in \cite{Horowitz:2000kx} to approximate localised Schwarzschild black holes in the $AdS_p \times S^q$ solutions of string and M-theory.

As we are interested in finding solutions corresponding to localised D3 branes at the tip of this flux background, it will be useful to first recall the D3 brane solution:
\begin{align}
\label{eq:D3} 
ds^2 &= H^{-1/2} \left(-f dt^2 + dx_1^2 + dx_2^2 + dx_3^2 \right) + H^{1/2} \left( \frac{dr^2}{f} + r^2 d\Omega_5^2 \right), \\
\mathcal{F}_5 &= \varepsilon_B \coth\beta \, dH^{-1} \wedge dt \wedge dx^1 \wedge dx^2 \wedge dx^3, \nonumber \\
H &= 1 + \sinh^2 \beta \left(\frac{r_+}{r} \right)^4, \qquad f = 1 - \left( \frac{r_+}{r} \right)^4. \nonumber
\end{align}
As before, we will take $\beta \ge 0$ and allow $\varepsilon_B = \pm 1$ to characterise the difference between positive and negative charge. The extremal limit is $\beta \rightarrow \infty$ while keeping $r_0^2 \equiv r_+^2 \sinh\beta$ fixed. The case $\beta = 0$ corresponds to no charge at all, in which case the solution is simply $\text{Schw}_7 \times \mathbb{R}^3$. A useful coordinatization of the 5-sphere is 
\begin{equation}
d\Omega_5^2 = d\psi^2 + \sin^2\psi d\Omega_2^2 + \cos^2\psi d\widetilde{\Omega}_2^2,
\end{equation}
which connects to the coordinates employed in the toy flux background \eqref{eq:IIBflux} upon identifying $r_1 = r \sin\psi$, $r_2 = r \cos\psi$.

We now wish to consider a linear $G_3$ perturbation of the D3 brane solution. Because we wish to consider localised branes in the flux background, we will require the $G_3$ flux to approach the flux background form asymptotically:
\begin{align} 
\label{eq:asymptoticG3flux}
\lim_{r\rightarrow \infty} G_3 & = m \left( r_2^2 dr_2 \wedge d\widetilde{\Omega}_2 - i \varepsilon_F r_1^2 dr_1 \wedge d\Omega_2 \right), \\
& = m r^2 \left( \cos^2\psi \left( \cos \psi dr - r \sin\psi d\psi \right) \wedge d\widetilde{\Omega}_2 - i \varepsilon_F \sin^2\psi \left( \sin\psi dr + r\cos\psi d\psi \right)\wedge d\Omega_2 \right). \nonumber 
\end{align}
Using this asymptotic form as a guide, a suitable ansatz for the perturbation is
\begin{align} 
G_3 &= m r^2 \cos^2\psi \left( g_1\cos \psi dr - g_2 r \sin\psi d\psi \right) \wedge d\widetilde{\Omega}_2 \\
& - i \varepsilon_F m r^2 \sin^2\psi \left( g_3 \sin\psi dr + g_4 r\cos\psi d\psi \right)\wedge d\Omega_2. \nonumber
\end{align}
As in \cite{Horowitz:2000kx}, we find that the perturbation functions can be taken to be simply functions of $r$ only. The Bianchi constraint $dG_3 =0$ is satisfied for
\[ g_2 = g_4, \qquad g_1 = g_3 = \frac{1}{3} r^{-2} \partial_r \left( r^3 g_4 \right). \]
The linearised equation of motion $d \star G_3 + i G_3 \wedge F_5 =0$ then reduces to simple ODE \footnote{A similar equation was derived in \cite{Freedman:2000xb} in the context of studying finite temperature effects in Polchinski-Strassler polarisation. They were not concerned with matching the solution to a flux background, and therefore dropped the asymptotic region (the overall constant in $H$).}
\begin{equation}
\label{eq:IIBperteq}
r^2 f H g_4'' + \left(r^2 H f'-r^2 f H'+7 r f H\right)g_4' + 3\left(H \left(r f'+3 f-3\right)+r H' (\varepsilon  \coth\beta-f)\right)g_4  = 0,
\end{equation}
where $'$ indicates derivative with respect to $r$, and we have introduced $\varepsilon = \varepsilon_F \varepsilon_B$. Note that the perturbation equation only depends on the relative sign of the charge of the brane and flux background. For $\varepsilon = 1$ they are aligned, and for $\varepsilon = -1$ they are anti-aligned. We will solve this equation separately for the non-extremal and extremal cases.

\subsubsection{Non-Extremal Case
\label{subsubsec:IIBlocalisedfinitetemp}}
\begin{figure}[H]
\begin{center}
\includegraphics[width=0.5\textwidth]{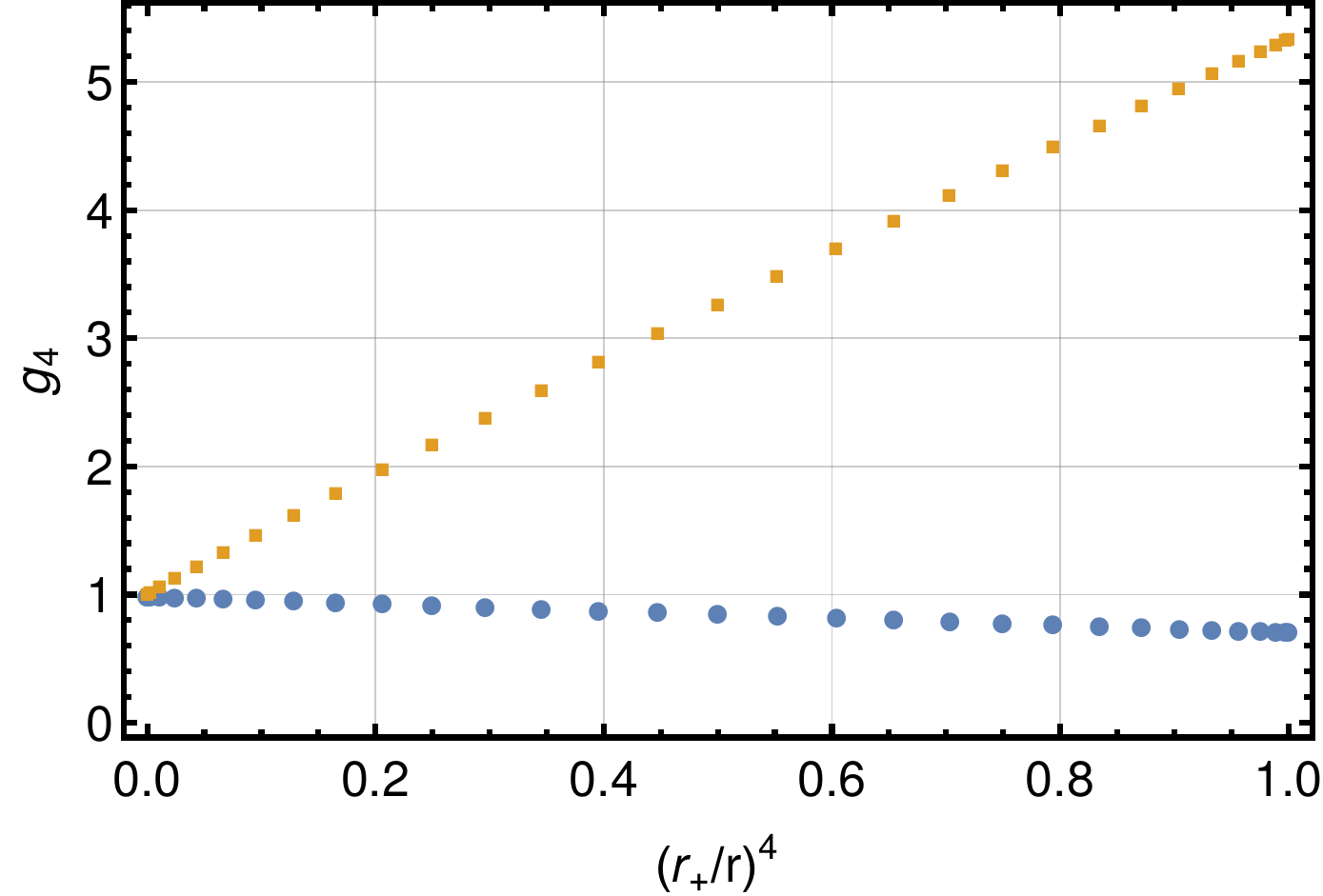}
\end{center}
\caption{The perturbation function $g_4$ for $\beta = 1$ and $\varepsilon = 1$ (lower curve) and $\varepsilon = -1$ (upper curve). Note that the flux at the horizon, $(r_+/r)^4 = 1$, is significantly larger for the anti-aligned $\varepsilon = -1$ case.
\label{fig:D3influx}}
\end{figure}

Unfortunately, we were unable to find analytic solutions for the general non-extremal case, and will therefore turn to numerical methods. To facilitate the numerical evaluation, convert to the compactified coordinate $x = (r_+/r)^4$, for which the equation becomes
\begin{equation}
\label{eq:IIBlocalisedfinitetemp}
\partial_x^2 g_4 + \left( \frac{1+x+x(3-x) \sinh^2\beta}{2x(x-1)\left(1+x\sinh^2\beta\right) } \right) \partial_x g_4 + \left( \frac{ 6\varepsilon\sinh2\beta -3 +3(3x-4) \sinh^2\beta}{16x(x-1)\left(1+x\sinh^2\beta\right)} \right) g_4 =0. 
\end{equation}
The boundary conditions we desire are that the perturbation approaches the form of the flux background at infinity $g_4(x=0) = 1$, and that the perturbation be regular at the horizon. By expanding $g_4$ in a power series around $x=1$, this last condition can be seen to be equivalent to
\begin{equation}
\left[ \frac{3}{16} \left(1-4 \,  \varepsilon \tanh\beta \right) g_4 - \partial_x g_4 \right]_{x=1}= 0.
\end{equation}
We solved \eqref{eq:IIBlocalisedfinitetemp} numerically, and in Fig.~\ref{fig:D3influx} a typical solution is plotted.

\begin{figure}[H]
\begin{center}
\includegraphics[width=0.5\textwidth]{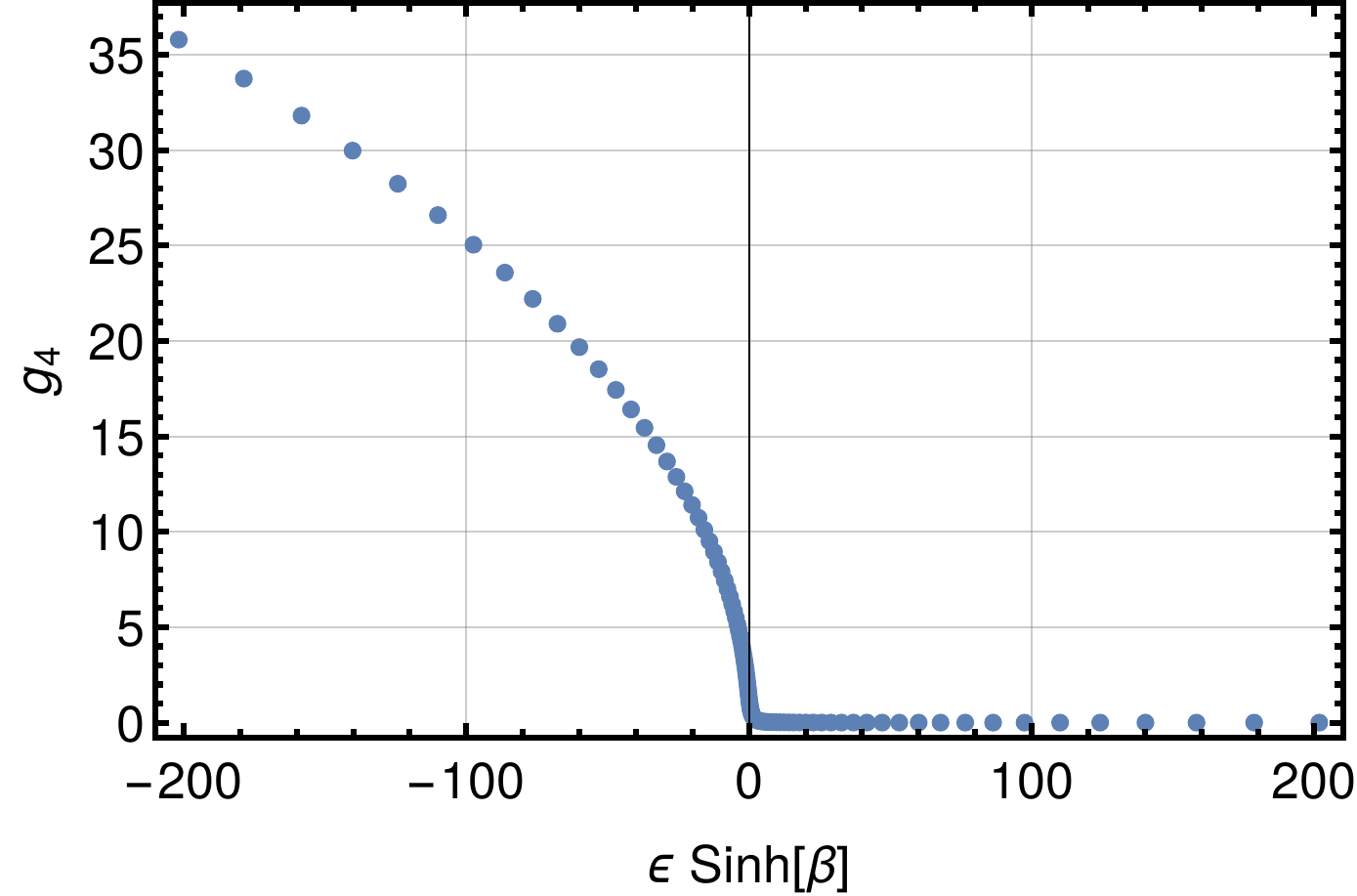}
\end{center}
\caption{The value of the rescaled perturbation $g_+$ at the horizon. As extremality is approached, the flux either vanishes $(\varepsilon = 1)$, or diverges $(\varepsilon = -1)$. Recall that our convention is that $\beta\ge 0$ always. Fitting the data indicates that $g_+(r_+) \sim e^{\beta/2}$ as $\beta \rightarrow \infty$ for $\varepsilon = -1$, which corresponds to $g_4(r_+) \sim e^{2\beta}$.
\label{fig:q1athorizonD3}}
\end{figure}

We find, remarkably, that a solution exists \textit{for both signs of the charge!} This is the central result of this paper. When the charges are anti-aligned, the flux at the horizon is significantly larger than the case when they are aligned, as might have been expected. The norm of the flux at the horizon is 
\begin{equation}
|G_3|^2 = 6m^2 \frac{g_4(r_+)^2}{\cosh^3\beta}.
\end{equation}
In Fig.~\ref{fig:q1athorizonD3} the quantity $g_+ \equiv g_4(r_+)/\cosh^{3/2}\beta \propto |G_3|$ is plotted as a function of $\beta$ for both signs of $\varepsilon$. For large positive charge, the flux at the horizon vanishes, while for large negative charge it grows without bound. 

Next, we are obliged to conduct a few checks on our calculation and to compare it with previous results. Firstly, an analytic solution can be found for the special case of $\beta =0$:
\begin{equation} 
g_4(x) = \, _2F_1\left(-\frac{3}{4},\frac{1}{4};-\frac{1}{2};x\right) - \left( \frac{2  \Gamma \left(\frac{3}{4}\right) \Gamma \left(\frac{7}{4}\right)}{\Gamma \left(\frac{1}{4}\right)^2} \right) x^{3/2} \, _2F_1\left(\frac{3}{4},\frac{7}{4};\frac{5}{2};x\right). 
\end{equation}
Comparing this analytic expression against the numerical solution provides a useful check on the numerics which is passed excellently.

Secondly, as in \cite{Horowitz:2000kx}, even though a static and non-singular perturbation has been found, because the flux does not vanish at the horizon it could be the case that stress-energy is flowing across the horizon and that the solution is not consistent with the Raychaudhuri equation, which requires that the energy flux across a static horizon must vanish. Letting $k^a = (\partial/\partial t)^a$ denote the null generators of the horizon, it is easy to verify that $T_{ab} k^a k^b = 0$, as $F_5$ is the only flux with $t$-legs, and is uncorrected at this order: $T_{ab} k^a k^b \propto (F_5)_{a cdef} (F_5)_{b}{}^{cdef} k^a k^b = 0$.

As mentioned in the Introduction, this result appears to violate the recent no-go theorem of \cite{Blaback:2014tfa}, and therefore some comments are in order. This theorem has a number of loop-holes. They are: (a) it could be that the boundary term does not satisfy their Eq.~4.24, (b) that $C(M_0,T)$ is not analytic \footnote{The definition of the boundary term and $C(M_0,T)$ can be found in \cite{Blaback:2014tfa}.}, and (c) it could be the case that singular terms conspire to cancel. Unfortunately, due to our approximation scheme wherein the solution is only known near the branes, we cannot evaluate options (a) and (b). We will however argue that option (c) is the relevant loop-hole. 

Recall the perturbation of the smeared black brane \eqref{eq:IIBsmearedpert}. The symmetries of the problem result in a single integration constant, which for general values lead to a singular flux $|H_3|^2 \sim f^{-1}$. Only for a particular choice of the constant do different singular terms cancel to make a regular perturbation. This cancellation of singular terms is actually standard practice in constructing hairy black hole solutions--unless the boundary conditions are chosen to make the hair regular at the horizon, the generic solution will diverge there. In the smeared case, since there is only one constant, the flux cannot be chosen to both be regular and to have the opposite orientation as the brane. In the localised case, there are fewer symmetries and therefore the flux has enough freedom to interpolate between a smooth value at the horizon and the desired value at infinity. Having two integration constants now allows both for the cancellation of singular terms at the horizon as well as for the flux to asymptote to either an ISD or AISD form far from the brane. It appears that this is how the no-go theorem is evaded.


The localised solution constructed here has important implications for the existence of the KPV metastable state. According to the Gubser criterion the existence of a localised black brane with negative charge implies that the singularity associated with anti-D3 branes in the KS solution are physical, and can in fact be resolved by string theory. As the brane approaches the negatively charged extremal limit, the flux diverges outside the horizon, which suggests a resolution by polarisation. We will examine this possibility next.

\subsubsection{Extremal Case
\label{subsubsec:IIBextremal}}
We next turn to the extremal case, for which the perturbation equation \eqref{eq:IIBperteq} has a very simple analytic solution:
\begin{align}
\label{eq:IIBextremalpert} 
(\varepsilon = 1) & \qquad g_4 = c_0 + c_1 \left(\frac{r_0^6}{r^6} + \frac{3}{5} \frac{r_0^{10}}{r^{10}} \right), \\
(\varepsilon = -1) & \qquad g_4 = c_0 \left(1 + 3 \frac{r_0^4}{r^4} \right) + c_1 \frac{r_0^6}{r^6}, \nonumber
\end{align}
where $c_0$, $c_1$ are integration constants. The boundary condition that $g_4(\infty) = 1$ requires $c_0 = 1$. Generically, the perturbation is singular in the IR. The norms of the 3-form fluxes are
\begin{align} 
(\varepsilon = 1) \qquad |F_3|^2 \sim |H_3|^2  \sim r^{-14}, \quad c_1 \neq 0, \qquad |F_3|^2 \sim |H_3|^2 \sim r^6, \quad c_1 = 0, \\
(\varepsilon = -1) \qquad |F_3|^2 \sim |H_3|^2 \sim r^{-6}, \quad c_1 \neq 0, \qquad |F_3|^2 \sim |H_3|^2 \sim r^{-2}, \quad c_1 = 0. \nonumber
\end{align}
For $\varepsilon = 1$, the norm can be made non-singular at the brane location by setting $c_1 = 0$. In fact, for this case an exact solution exists at the non-linear level corresponding to extremal branes added to the flux background. This just corresponds to adding a singular harmonic source term to the warp factor $H$, just as in the previous section. Explicitly, the solution fits the same ansatz as the flux background \eqref{eq:IIBflux}, but the warp factor is now \footnote{Note that the branes added here are localised. The addition of D3 branes smeared over 3 dimensions corresponds to the addition of a term $a_0/r_1 = a_0/(r \sin\psi)$, as discussed previously.}
\begin{equation}
H = 1 + \frac{r_0^4}{r^4} - \frac{m^2 r^2 \sin^2\psi }{6}.
\end{equation}
For $\varepsilon = -1$, there is no way to keep the asymptotic condition $c_0 = 1$ and to also keep the norm non-divergent. As expected from the extremal limit of the previous section, the fluxes must diverge as $r \rightarrow 0$. The interpretation of this singularity is of crucial importance for understanding the problem of anti-branes in flux backgrounds. That it can be shielded by a finite-temperature horizon implies that it can be resolved by string theory.

The obvious candidate resolution mechanism is brane polarisation. In fact, the set-up is very similar to the one considered by Polchinski-Strassler in the supersymmetric case \cite{Polchinski:2000uf}. The anti-D3 branes source an $AdS_5$ throat, into which the singular $G_3$ flux will leak. If the fluxes take the right form, polarisation will occur and the singularity will be resolved in a manner similar to the way in which  Polchinski-Strassler resolves the singularity associated with a mass deformation of the $\mathcal{N}=1^*$ theory, which we now review very briefly.

Taking both the near-brane and extremal limit of D3 brane solution \eqref{eq:D3} results in the line element
\begin{equation} 
ds^2 \approx \Big( \frac{r}{r_0} \Big)^2 dx_{\mu}dx^{\mu} + \Big( \frac{r_0}{r} \Big)^2 dr^2 + r_0^2 d\Omega_5^2,
\end{equation}
which is simply $AdS_5 \times S^5$. Polchinski and Strassler considered Lorentz invarance-preserving $G_3$ perturbations of this solution. The perturbation transforms under the transverse $SO(6)$ according to two possible representations: the $\textbf{10}$ and the $\overline{\textbf{10}}$. There are two solutions for each representation, each of which has a definite scaling with $r$, i.e. for $r \rightarrow \lambda r$, $G_3 \sim \lambda^{-\Delta} G_3$. For this very symmetric case of perturbations of $AdS_5$, only a single power of $r$ is present, as opposed to the more general case where a whole power series would be expected. The possible conformal dimensions are
\begin{align}
\textbf{10}:& \qquad \Delta = 7, \qquad \Delta = -3, \\ 
\overline{\textbf{10}}:& \qquad \Delta = 3, \qquad \Delta = 1 \nonumber. 
\end{align}
For $\Delta \le 0$ the mode is regular in the IR $(r\rightarrow 0)$, and for $\Delta > 0$ it diverges. Polchinski and Strassler studied the perturbation corresponding to the $\Delta = 1$ mode in the supersymmetric case and found that the singularity was cured through the physics of brane polarisation. 

Returning now to the case of $G_3$ perturbations of the full D3 brane solution, as opposed to simply the near-horizon $AdS_5 \times S^5$, we see that for the case $\varepsilon = -1$, there are three modes present, that is, $G_3$ is a sum of three terms with distinct definite scalings:
\begin{equation}
G_3 \sim c_1 \lambda^{-3} + c_0 \left( \lambda^{-1} + \lambda^{3} \right).
\end{equation}
The perturbation therefore contains the conformal dimensions $\Delta \in (3,1,-3)$. Taking $c_0 \neq 0$, as required by the boundary condition that the flux approach \eqref{eq:asymptoticG3flux} asymptotically forces the modes $\Delta = 1,-3$ to be turned on. The most singular $\Delta = 3$ mode can be eliminated by setting $c_1 = 0$ . What remains is the non-singular $\Delta = -3$ mode, which becomes vanishingly weak and therefore negligible in the deep IR ($r \rightarrow 0$), and the $\Delta = 1$ mode, which diverges down the throat \cite{Polchinski:2000uf}. \footnote{There is a nice connection with the weaker divergence associated with setting $c_1 = 0$ for the $\varepsilon = -1$ case, and the no-go results of \cite{Blaback:2014tfa} which in the present zero temperature case imply that the 3-form fluxes must diverge as $|H_3|^2 \sim H^{1/2} \sim r^{-2}$. Unlike the finite temperature case, this divergence cannot be cured by adjusting constants to procure cancellations.} 

In the original Polchinski-Strassler analysis, the form of the $\Delta = 1$ mode was chosen to be supersymmetric, which allowed the polarisation potential to be determined from simply the linear $G_3$ flux perturbation. The present case is more complicated, as the form of the divergent $\Delta = 1$ mode can be shown to not preserve supersymmetry. In the $SO(3)$-invariant, non-supersymmetric version of Polchinski-Strassler \cite{Zamora:2000ha}, the polarisation potential depends on 3 parameters, $m_{\text{PS}},m'_{\text{PS}},\mu_{\text{PS}}$, which are defined as follows. The $\Delta = 1$ $G_3$ perturbation can be written as 
\begin{equation}
G_3 \propto \frac{1}{r^4} \left(T_3 - \frac{4}{3} V_3 \right),
\end{equation}
where $T_3$ and $V_3$ are 3-forms on the $\mathbb{R}^6$ transverse to the branes defined in \cite{Polchinski:2000uf} and reproduced in Appendix \ref{sec:G3complex}. The coefficient of the $(1,2)$ component of $T_3$ is denoted $m_{\text{PS}}$, while $m'_{\text{PS}}$ is the coefficient of the $(3,0)$ component. $\mu_{\text{PS}}$ is a parameter that comes from the second order backreaction of the $G_3$ perturbation. In the supersymmetric case, $m'_{\text{PS}}=\mu_{\text{PS}}=0$, and the linear flux perturbation suffices to compute the polarisation potential. In the present case, the form of the $\Delta = 1$ mode of the flux is shown in Appendix~\ref{sec:G3complex} to give rise to $m_{\text{PS}}' = m_{\text{PS}}$, and the perturbation therefore breaks all supersymmetry. In order to definitively determine whether or not polarisation resolves the singularity, the parameter $\mu_{\text{PS}}$ would also need to be computed, which would require going beyond the linear approximation considered here.

\begin{figure}[]
\begin{center}
\includegraphics[width=0.7\textwidth]{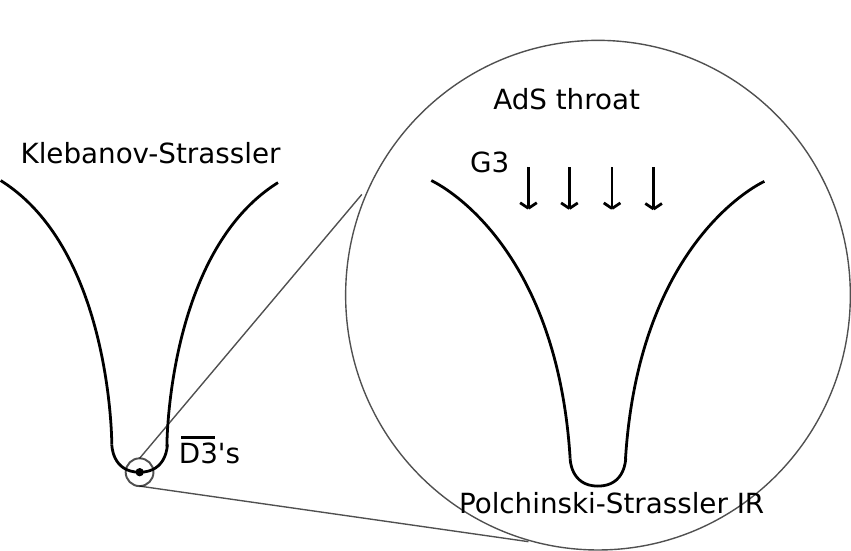}
\end{center}
\caption{Schematic depiction of the proposed supergravity solution corresponding to anti-D3 branes in the Klebanov-Strassler solution. The solution interpolates between a Klebanov-Strassler throat asymptotically, and a mass-deformed $AdS_5 \times S^5$ throat. The $\Delta =1$ mode of the $G_3$ flux grows down the throat, until it eventually induces polarisation, leading to the (as yet un-constructed) backreacted non-supersymmetric Polchinski-Strassler solution. It should be noted that each throat has a distinct radial coordinate. 
\label{fig:PSglueKS}
}
\end{figure}

Despite the lack of a definitive polarisation calculation, the finite-temperature and extremal cases together paint a compelling picture. In the extremal case of anti-branes added to the flux background, the anti-D3's source an $AdS_5$ throat into which the $G_3$ fluxes leak. It could have turned out that the most singular mode in the IR, the $\Delta = 3$ mode, was forced to be present from the requirement that the flux approach the flux background solution asymptotically, but that did not turn out the be the case. Instead, the only singular mode that leaks into the throat is the $\Delta = 1$ mode. As $r\rightarrow 0$ is approached, this mode grows until polarisation via the Myers effect likely takes over, leading to a completely non-singular, puffed-up brane configuration. The full solution corresponding to anti-branes in KS would then correspond to a gluing of the backreacted non-supersymmetric Polchinski-Strassler solution (which has yet to be explicitly constructed) to the KS geometry. This is schematically depicted in Fig.~\ref{fig:PSglueKS}. 

Our calculation in the extremal case is similar to one that appears in \cite{DeWolfe:2004qx}. The existence of negatively charged black branes at finite temperature therefore strengthens the picture considered there and in \cite{Dymarsky:2011pm} wherein the gravity dual of the KPV metastable state was argued to be a gluing of the polarised Polchinski-Strassler solution with the Klebanov-Strassler one. Finite temperature is seen to act as an IR regulator of the un-polarised singular solution. For negatively charged solutions close to extremality, one would expect that these black branes are unstable to a finite temperature version of brane polarisation as the flux is very large outside the horizon. \footnote{Finite temperature effects in the $\mathcal{N}=1^*$ theory are treated in \cite{Freedman:2000xb}.}

The existence of negatively charged black branes at finite temperature provides evidence, according to the Gubser criterion, that the anti-brane singularity is resolved. The diverging form of the flux in the extremal limit strongly suggests that  the appropriate resolution mechanism is brane polarisation, a conclusion at odds with previous calculations which suggested that polarisation only occurs in certain channels, and that the throat is unstable to fragmentation \cite{Bena:2012tx, Bena:2012vz, Bena:2014bxa, Bena:2014jaa}. Note that there is no indication of a fragmentation instability of the blackened anti-D3 brane solutions constructed here. In these works the the potentials were computed for probe branes in the solutions corresponding to \textit{smeared} anti-branes in flux backgrounds, and it was then argued that the potential for localised anti-branes could be extracted from the smeared result. Central to the argument is the remarkable fact that in Polchinksi-Strassler the polarisation potential for branes in a mass-deformed $AdS_5$ throat is independent of the warp factor $H$. Since the potential is independent of $H$, the potential for a probe brane in the mass-deformed throat created by a stack of D3 branes is the same as that created by a configuration of D3-branes smeared over an $S^2$, a key point in the original PS analysis. 

However, the smearing in the present case is much more drastic than the smearing considered in PS. For the case of anti-D3 branes added to the toy flux background \eqref{eq:IIBflux}, it is clear that the smearing not only alters the form of the near-brane $AdS_5$ throat, but also greatly restricts the form of the fluxes in the throat. In particular, it is no longer appropriate to consider general $SO(6)$ representations of $G_3$ perturbations, as these will generically break the symmetry imposed by smearing. It is important to note that the asymptotic form of the flux is invariant under the symmetries imposed by smearing, but the form of the flux down the throat, where polarisation would occur, is not. This fact, together with the finite temperature case where the smearing clearly constrains the form of the flux to be such that it is unable to be both regular at the horizon and asymptote to either self-duality signature, strongly suggests that smearing destroys the resolution mechanism.

The calculations presented here indicate that the singularities associated with anti-D3 branes at the tip of the Klebanov-Strassler throat are resolved and that the KPV metastable state exists. In Appendix \ref{sec:M-theory} we repeat the above analysis for a toy flux background of M-theory which approximates the IR of the CGLP solution \cite{Cvetic:2000db}. All the main results carry over, and we therefore turn to a concluding discussion.

\section{Conclusion
\label{sec:conclusion}}
In this paper we studied localised black branes in toy flux backgrounds in both string theory and M-theory. The finite temperature solution was constructed to leading order in an approximation scheme where the flux was treated perturbatively near the branes. Remarkably, black brane solutions were found to exist for either sign charge, indicating that the singularity associated with localised anti-branes can be shielded behind a smooth horizon. According to the Gubser criterion, the anti-brane singularities are then physical and resolvable by string theory.

As the charge of the black brane is taken taken to be large and negative, the flux at the horizon is seen to increase without bound. The natural resolution mechanism is therefore brane polarisation \`{a} la Polchinski-Strassler \cite{Polchinski:2000uf} or its M-theory analogue \cite{Bena:2000zb}. And indeed, when the extremal anti-brane limit is taken, the divergent flux is of the form known to be resolved by polarisation in the supersymmetric case. Because the matching of the $AdS$ throat sourced by the branes with the flux background breaks supersymmetry, a non-supersymmetric version of Polchinski-Strassler \cite{Zamora:2000ha} or its M-theory analogue is required. In order to determine the polarisation potential in the non-supersymmetric case, and to definitively address whether or not polarisation resolves the singularity, the solution will need to be extended to second order in the flux perturbation. If polarisation is indeed the correct resolution mechanism, which we view as likely, then the metastable non-SUSY states corresponding to puffed-up branes not only exist, but also possess supergravity descriptions. In the Type IIB case this corresponds to a gluing of the polarised Polchinski-Strassler solution to a Klebanov-Strassler throat.

The calculations of the current work are in tension with a number of recent results. The existence of localised black holes in the IIB flux background we considered appeared to be ruled out by the no-go theorem of \cite{Blaback:2014tfa}, however we argued that this theorem is evaded by a loop-hole. The fluxes are indeed generically singular at the horizon, but by a choice of integration constants these can be made regular in both the smeared and localised cases. The symmetry imposed by smearing is rather restrictive, and does not allow the flux at the horizon to be both regular and anti-aligned with the asymptotic value. In the less symmetric localised case, there is enough freedom to simultaneously enforce regularity at the horizon as well as to fix the orientation to be either sign asymptotically. 

Additionally, previous calculations used smeared results to argue that polarisation only occurs in certain channels for the localised case, and that the $AdS$ throats were unstable to fragmentation instabilities (the so-called tachyons), thus rendering the anti-brane singularities unphysical \cite{Bena:2012tx, Bena:2012vz, Bena:2014bxa, Bena:2014jaa}. However our calculation of the flux near the brane in both the extremal and non-extremal cases strongly suggests that polarisation does occur, and yields no indication of the fragmentation instability. We argued that the discrepancy could be due to the restrictive symmetry imposed by smearing. It appears that smearing destroys the resolution mechanism, a possibility in accord with existence of negatively charged anti-branes in the localised case, and the absence of them in the smeared case. If polarisation is not the resolution mechanism, then there is either some novel mechanism yet to be discovered which cures the singularity, or else the Gubser criterion as it is currently understood fails and requires revision.

Despite recent work suggesting an increasingly dire outlook for the fate of anti-branes in flux backgrounds and their many interesting applications, there are now a few distinct and complementary results which suggest that the metastable states do in fact exist as originally understood. First, there are the original polarisation calculations \cite{Kachru:2002gs, Klebanov:2010qs}. Secondly, in \cite{Michel:2014lva} it was convincingly argued that a single brane should be non-singular and described by effective field theory. The calculations of the present work indicate that the physics of many branes is also non-singular (after they polarise, or are resolved by some new mechanism). There is no need to resort to time dependence to resolve the singularity \cite{Blaback:2011pn, Blaback:2012nf, Danielsson:2014yga}, and there are no challenges to the de Sitter landscape or other applications of the non-SUSY metastable states.

The results of this paper suggest many avenues for future work. It is of obvious interest to go beyond the linear approximation considered here, and to construct the solutions in a systematic matched asymptotic expansion for both the extremal and non-extremal cases. In the extremal case, this would allow for a definitive answer to the question of whether polarisation resolves the singularities and results in a smooth, stable throat. It would also be very worthwhile to explore the solution space of black branes in flux backgrounds further. In addition to localised and smeared solutions, there are likely to exist inhomogeneous black branes as well. In many ways, the problem of black brane solutions in flux backgrounds seems qualitatively  similar to the problem of black hole/string solutions in 5d Kaluza-Klein gravity.\footnote{For a review, see \cite{Horowitz:2011cq}. Black holes in a simpler family of flux backgrounds were studied recently in \cite{Hartnett:2014zca}.} In particular, it seems likely that there is a very interesting relationship between inhomogeneities of the black branes and charge. Additionally, the negatively charged localised black holes constructed here might themselves be unstable to a finite-temperature version of Polchinski-Strassler \cite{Freedman:2000xb}. In a future work we plan to explore these and related issues.

\begin{acknowledgments}
Some of the ideas contained in the present work originated in a collaboration with G. Horowitz and A. Puhm, and a separate collaboration with \'{O}. Dias and J. Santos, and I am very grateful to them for contributing to our understanding of the problem and for their encouragement throughout this project. I also wish to thank J. Brugger, D. Marolf, B. Michel, E. Mintun, and especially J. Polchinski for useful discussions, as well as the authors of \cite{Michel:2014lva} for sharing early drafts of their paper. Lastly, we wish to thank \'{O}. Dias, S. Fischetti, G. Horowitz, W. Kelly and J. Santos for discussions and their comments on a draft of this paper. This work was supported by NSF grant PHY12-05500, and by funds provided by the University of California, Santa Barbara.
\end{acknowledgments}

\begin{appendix}
\section{Strings are Well Behaved in the Toy Flux Backgrounds}
{\label{sec:naked}}
In this section we show that although the toy flux backgrounds \eqref{eq:IIBflux}, \eqref{eq:Mtheoryflux} are certainly singular at the locus $H = 0$, strings propagating on these backgrounds are actually non-singular and completely well behaved. The work presented in this appendix is a result of joint work with G. Horowitz and A. Puhm.

Both the toy flux backgrounds considered in this paper and the KS and CGLP solutions are actually U-dual to plane-fronted waves with parallel propagation, or pp-waves for short \cite{Cvetic:2002hi}. PP-wave spacetimes are of the form
\begin{equation}
ds^2 = -2 du dv + F(u,x_{\perp})du^2 + dx_{\perp}^2, 
\end{equation}
where $dx_{\perp}^2$ is some Ricci-flat  transverse space. The vector $\ell_{\mu} = \partial_{\mu} u$ is null and covariantly constant, and both the Riemann and Ricci curvature tensors can easily be written in terms of $\ell$ and $F$ \cite{Horowitz:1990sr}:
\begin{equation}
R_{abcd} = 2\ell_{[a} \partial_{b]} \partial_{[c} F \ell_{d]}, \qquad R_{ab} = -\frac{1}{2} \left( \nabla^2_{\perp} F \right) \ell_a \ell_b,
\end{equation}
where $\nabla_{\perp}^2$ is the scalar Laplacian on the transverse space. Note that the $u$-dependence is completely arbitrary.

Because the curvature is null, all curvature invariants of pp-waves vanish, in particular the Ricci scalar. When $F$ depends quadratically on the transverse coordinates, the spacetime is said to be an exact plane wave. If $\nabla_{\perp}^2 F = 0$, then  the plane wave is a vacuum solution, otherwise the geometry will be sourced by a stress tensor that is also purely null, $R_{ab} = T_{ab}$.

For concreteness, consider the Type IIB toy flux background \eqref{eq:IIBflux}. although similar comments apply to the other flux backgrounds considered here (the toy M-theory solution considered in Appendix \ref{sec:M-theory}, and the KS and CGLP solutions). This can be dualised to a Type IIB pp-wave according to the following chain of transformations \footnote{An alternative duality chain involving $T_{x_3} T_{x_2} S T_{x_1}$ results in a IIA pp-wave background.}
\begin{equation} \left(\begin{array}{c} D3\\ D5\end{array}\right) \underrightarrow{T_{x_3}} 
\left(\begin{array}{c} D2\\ D4\end{array}\right) \underrightarrow{\uparrow x_{11}}
\left(\begin{array}{c} M2\\ M5\end{array}\right) \underrightarrow{\downarrow x_2}
\left(\begin{array}{c} F1\\ D4\end{array}\right) \underrightarrow{T_{x_1}}
\left(\begin{array}{c} P\\ D3\end{array}\right), \label{KSchain}
\end{equation} 
where $T_{x_i}$ indicates T-duality on the $x_i$-coordinate, and $\updownarrow x_i$ indicates either the M-theory uplift or reduction along $x_i$. The D3 charge dissolved in flux of the original solution becomes momentum while the fractional D3 brane charge (D5 branes wrapping additional two-cycles) becomes D3 charge. The resulting background is
\begin{equation}
d{s'}^{2}_{10} = 2\varepsilon_F  dt dx_1 + H dx_1^2 + \left( dx_3^2 +dx_{11}^2 + ds_6^2 \right),
\label{eq:KSppmetric}
\end{equation}
where $ds_6^2$ is the 6-dimensional transverse space. The warp factor $H$ takes the place of $F(u,x_{\perp})$, and depends only on $r_1$ The last coordinate to be T-dualised upon, $x_1$, serves as the null coordinate. The Ricci tensor is
\begin{equation}
R_{ab} = - \frac{1}{2} \left( \nabla_6^2 H \right) \partial_a x_1 \partial_b x_1.
\end{equation}
Since $\nabla_6^2 H \neq 0$, this geometry must be supported by matter. The dilaton is constant, $e^{2\Phi'}=1$, and the only non-vanishing form field is
\begin{equation}
F_5' = [F_3^0 \wedge dx_3 + H_3^0 \wedge dx_{11}]\wedge dx_1,
\end{equation}
where $F_3^0$ and $H_3^0$ denote the original three-form fluxes of the toy flux background. The five-form flux is null and self-dual. The stress tensor is then
\begin{equation} 
T_{ab} = \frac{1}{2} m^2 \partial_a x_1 \partial_b x_1. 
\end{equation} 
An interesting feature of this duality frame is that the imaginary self-duality property of the original 3-form flux corresponds to the fact that the curvature and matter are null.

We now use the above duality chain to show that strings are perfectly well behaved in the toy flux background solution, despite the appearance of a singularity at the locus $H=0$. The original solutions' naked singularities can be studied using the dual plane wave solutions. The physics of strings propagating on the original solutions is dual to the physics of strings propagating on the dual plane waves, since the worldsheet theory of the strings is duality-invariant.

Recall that plane waves are a special case of pp-waves in which the $F$ function is quadratic in the $x_{\perp}$ coordinates. The warp factor of the toy flux background is $H = 1 - m^2 r_1^2/6$, and so the dual solution is an exact plane wave. The fact that these simple flux backgrounds dualise to plane waves is particularly nice because plane waves have the special property in string theory that strings may be quantized exactly \cite{Horowitz:1990sr} \footnote{More precisely, because the curvatures are null, pp-waves are solutions of the supergravity equations to any order in $\alpha'$. Plane waves, moreover, are solutions even non-perturbatively in $\alpha'$. Additionally, pp-waves are the only non-flat spacetime to admit lightcone gauge on the string worldsheet, and for the special case of plane waves, the string equations of motion are linear and different modes decouple, allowing the string to be quantized exactly.} After dualising, the plane wave is completely non-singular at the locus $H=0$. To see this, note the radius at which $H=0$ is no longer gauge invariant; under the coordinate transformation $t=\tilde{t} + D x^1$, with $D$ a constant, the radius at which $H=0$ is shifted. Therefore, whereas $H=0$ in the original solutions corresponded to a naked singularity, the geometry is completely regular in the dual plane waves. Strings quantized on the plane wave backgrounds would not experience any singular behaviour, and therefore we can conclude that neither would they experience singular behaviour in the original flux background solutions. 

\section{Explicit Form of the $G_3$ Flux in Complex Coordinates}
{\label{sec:G3complex}}
In this appendix we verify that the form of the $G_3$ perturbation considered in Sec.~\ref{subsubsec:IIBextremal} breaks the supersymmetry of the $AdS_5$ throat with $m'_{\text{PS}} /m_{\text{PS}} = 1$. 

First, introduce the complex coordinates $z^i = x^i + i y^i$, for $i=1,2,3$. The metric on $\mathbb{R}^6$ is then
\begin{equation}
ds^2_{\mathbb{R}^6} = \sum_{i=1}^3 dz^i d\bar{z}^i = \sum_{i=1}^3 \left( dx_i^2 + dy_i^2 \right).
\end{equation}
These coordinates are related to the ones used in the perturbation of the $G_3$ flux for the localised case considered in Sec.~\ref{subsec:IIBlocalised} by
\begin{equation}
\sum_{i=1}^3 dx_i^2 = dr_1^2 + r_1^2 d\Omega_2^2
, \qquad \sum_{i=1}^3 dy_i^2 = dr_2^2 + r_2^2 d\widetilde{\Omega}_2^2,
\end{equation}
with $r_1 = r \sin\psi$, $r_2 = r \cos \psi$. In the anti-aligned case $\varepsilon = -1$, the flux profile was found in \eqref{eq:IIBextremalpert} to be
\begin{equation}
g_4 = c_0 \left(1 + 3 \frac{r_0^4}{r^4} \right) + c_1 \frac{r_0^6}{r^6}.
\end{equation}
The most singular term in the IR ($r\rightarrow 0$) can be eliminated by setting $c_1 = 0$, and the asymptotic boundary condition requires $c_0=1$. For this choice, the leading term of the flux perturbation in the IR is
\begin{align} 
\label{eq:G3fluxIR}
G_3 \sim m \left(\frac{r_0^4}{r^2}\right) \Big( & \cos^2\psi \left(-\cos \psi dr - 3 r \sin\psi d\psi \right) \wedge d\widetilde{\Omega}_2 \\
& + i \sin^2\psi \left( -\sin\psi dr + 3 r\cos\psi d\psi \right)\wedge d\Omega_2 \Big) + ... \nonumber
\end{align}
In the notation of Polchinski-Strassler \cite{Polchinski:2000uf}, the $\Delta = 1$ flux perturbation is written as $G_3 \propto r^{-4} \left(T_3 - 4 V_3/3 \right)$, where in the $SO(3)$-invariant case (in which the chiral superfields have the same mass),
\begin{align}
T_3 &= m_{\text{PS}} \left(dz^1 \wedge d\bar{z}^2 \wedge d\bar{z}^3 + d\bar{z}^1 \wedge dz^2 \wedge d\bar{z}^3 + d\bar{z}^1 \wedge d\bar{z}^2 \wedge dz^3 \right) + m'_{\text{PS}} dz^1 \wedge dz^2 \wedge dz^3, \nonumber \\
V_3 &= \frac{x^q}{r^2} \left( x^m T_{qnp} + x^n T_{mqp} + x^p T_{mnq} \right).
\end{align}
The parameter $m_{\text{PS}}$ is therefore the coefficient of the $(1,2)$ component of $T_3$ and $m_{\text{PS}}'$ is the coefficient of the $(3,0)$ component. For $m_{\text{PS}}' \neq 0$, the perturbation breaks all supersymmetry of the $AdS_5$ throat. For the case at hand, \eqref{eq:G3fluxIR} can be shown to be equivalent to 
\begin{equation}
G_3 \sim - \frac{3i}{4} \frac{m}{m_{\text{PS}}} \left( \frac{r_0^4}{r^4} \right) \left(T_3 - \frac{4}{3} V_3 \right) + ... {}, \qquad m'_{\text{PS}} = m_{\text{PS}}.
\end{equation}
The overall normalization is irrelevant, and can be changed by simply rescaling $m$. This establishes that a non-supersymmetric version of the Polchinski-Strassler analysis is needed. 

\section{M2 Branes in M-theory Flux Backgrounds
\label{sec:M-theory}}
The analysis of Sec.~\ref{sec:IIB} can be repeated for M-theory. The M-theoretic version of the KS solution is the CGLP solution \cite{Cvetic:2000db}, which corresponds to fractional M2 branes at the tip of a higher dimensional version of the deformed conifold, known as the $n=3$ Stenzel space.\footnote{the $n=2$ and $n=1$ Stenzel spaces are the deformed conifold and the Eguchi-Hanson instanton, respectively.} The analogue of the KPV calculation was performed by  Klebanov and Pufu (KP) \cite{Klebanov:2010qs}, who found qualitatively similar results, namely that for small enough anti-M2 charge there is a metastable minima corresponding to anti-M2 branes polarised into M5 branes.

In this appendix we present a toy model of the CGLP solution. We then repeat the calculations of Sec.~\ref{sec:IIB} and are led to exactly the same conclusions--that localised M2 brane solutions exist for any sign charge in the non-extremal case. In the extremal case, resolution by polarisation again appears likely.

\subsection{A Toy Model of the CGLP Flux Background}
A toy model relevant for approximating the IR of CGLP is
\begin{align} 
\label{eq:Mtheoryflux}
ds^2 &= H^{-2/3} \left(-dt^2 + dx_1^2 + dx_2^2 \right) + H^{1/3}\left( dr_1^2 + r_1^2 d\Omega_3^2 + \sum_{i=1}^4 dy_i^2 \right), \\
G_4 &= \varepsilon_F \, dH^{-1} \wedge dt \wedge dx^1 \wedge dx^2 + m \alpha_4 , \nonumber
\end{align}
where 
\begin{equation}
\alpha_4 = dy^1 \wedge dy^2 \wedge dy^3 \wedge dy^4 - \varepsilon_F  r_1^3 dr_1 \wedge d\Omega_3 
\end{equation}
is the (anti)-self dual magnetic part of the four-form, $\star_6 \alpha_4 = -\varepsilon_F \alpha_4$, which should be thought of as analogous to the $G_3$ flux in the IIB case. The warp factor is
\begin{equation} 
H = 1 + \frac{a_0}{r^2} - \frac{m^2}{8} r_1^2, 
\end{equation} 
where, as in the IIB case, $a_0$ is proportional to the number of  smeared M2 branes which we take to be zero and $m$ controls the amount of flux. This solution is related via duality (reduction on $x_2$ followed by T-dualising the $x_1$ coordinate) to the plane wave solutions considered in \cite{Cvetic:2002hi}. This toy flux background also possesses a naked singularity, and the same comments as in the Type IIB case apply here as well (see Appendix \ref{sec:naked}).

The geometry near the tip of the $n=3$ Stenzel space consists of a finitely-sized $S^4$ and a shrinking $S^3$. In complete analogy with the discussion in Sec.~\ref{sec:IIB}, the flux background \eqref{eq:Mtheoryflux} is seen to capture the key features of the CGLP solution, with the main difference being that the transverse $S^4$ is replaced with a $\mathbb{R}^4$. For the purpose of discussing M2 branes localised at both tip of the background and also at a point in this transverse space, it is useful to introduce the coordinates 
\begin{equation}
\sum_{i=1}^4 dy_i^2 = dr_2^2 + r_2^2 d\widetilde{\Omega}_3^2, \qquad dy^1 \wedge dy^2 \wedge dy^3 \wedge dy^4 = r_2^3 dr_2 \wedge d\widetilde{\Omega}_3,
\end{equation}
where $d\widetilde{\Omega}_3^2$, $d\widetilde{\Omega}_3$ are the line elements and volume form for a second, distinct unit $S^3$. 

In complete analogy with Sec.~\ref{subsec:IIBpolarisation}, the Klebanov-Pufu polarisation calculation can be repeated here with the same conclusion--namely that the toy flux background always possess a minima corresponding to a puffed-up 5-brane wrapping an $S^3$ in the transverse $\mathbb{R}^4$ at the tip. This minima is the global minima, and there is no brane-flux annihilation process due to the non-compactness of the $\mathbb{R}^4$. For small anti-M2 charge, the potential is very similar to the KP potential, and the near-brane solution corresponding to localised anti-branes in the toy flux background should be identical to the analogous solution in the CGLP background.

\subsection{Localised Branes}
In this section we study localised branes in the toy M-theory flux background \eqref{eq:Mtheoryflux}. As  in Sec.~\ref{subsec:IIBlocalised}, our approach is very similar to that of Ref.~\cite{Horowitz:2000kx}. The thermal M2 brane solution is
\begin{align}
\label{eq:M2} 
ds^2 &= H^{-2/3} \left(-f dt^2 + dx_1^2 + dx_2^2 + \right) + H^{1/3} \left( \frac{dr^2}{f} + r^2 d\Omega_7^2 \right), \\
G_4 &= \varepsilon_B \coth\beta \, dH^{-1} \wedge dt \wedge dx^1 \wedge dx^2, \nonumber \\
H &= 1 + \sinh^2 \beta \left(\frac{r_+}{r} \right)^6, \qquad f = 1 - \left( \frac{r_+}{r} \right)^6, \nonumber
\end{align}
where as before $\beta \ge 0$ is the boost parameter and $\varepsilon_B = \pm 1$ characterises the difference between positive and negative charge. The extremal limit is $\beta \rightarrow \infty$ while keeping $r_0^3 \equiv r_+^3 \sinh\beta$ fixed. A useful coordinatization of the 7-sphere is 
\begin{equation}
d\Omega_7^2 = d\psi^2 + \sin^2\psi d\Omega_3^2 + \cos^2\psi d\widetilde{\Omega}_3^2,
\end{equation}
which connects to the coordinates employed in the toy flux background upon identifying $r_1 = r \sin\psi$, $r_2 = r \cos\psi$.

We now wish to consider a linear magnetic $G_4$ perturbation of the M2 brane solution. The $G_4$ flux should approach the \eqref{eq:Mtheoryflux} form asymptotically:
\begin{align} 
\lim_{r\rightarrow \infty} G_4^{(m)} & = m \left( r_2^3 dr_2 \wedge d\widetilde{\Omega}_3 -\varepsilon_F r_1^3 dr_1 \wedge d\Omega_3 \right), \\
& = m r^3 \left( \cos^3\psi \left( \cos \psi dr - r \sin\psi d\psi \right) \wedge d\widetilde{\Omega}_3  -\varepsilon_F \sin^3\psi \left( \sin\psi dr + r\cos\psi d\psi \right)\wedge d\Omega_3 \right) \nonumber. 
\end{align}
Using this asymptotic form as a guide, a useful ansatz for the perturbation is
\begin{align} 
\delta G_3 &= m r^3 \cos^3\psi \left(g_1  \cos \psi dr - g_2 r \sin\psi d\psi \right) \wedge d\widetilde{\Omega}_3 \\
& - \varepsilon_F m r^3 \sin^3\psi \left( g_3 \sin\psi dr + g_4 r\cos\psi d\psi \right)\wedge d\Omega_3 \nonumber,
\end{align}
where again the functions $g_i = g_i(r)$. The Bianchi constraint $dG_4 = 0$ is satisfied for
\[ g_2 = g_4, \qquad g_1 = g_3 = \frac{1}{4} r^{-3} \partial_r \left( r^4 g_4 \right), \]
and the linearised equation of motion $d \star G_4 + \frac{1}{2} G_4 \wedge G_4 =0$ then reduces to the ODE
\begin{equation}
r^2 f H g_4'' + \left(r^2 H f' - r^2 f H' + 9 r f H \right) g_4' + 4\left( H \left( r f' + 4 f - 4\right) + r H' \left( \varepsilon \coth\beta - f \right) \right) g_4 = 0,
\end{equation}
where again $\varepsilon = \varepsilon_F \varepsilon_B$. We will again consider the non-extremal and extremal cases separately.

\subsubsection{Non-Extremal Case \label{subsubsec:Mtheorylocalisedfinitetemp}}
\begin{figure}[H]
\begin{center}
\includegraphics[width=0.5\textwidth]{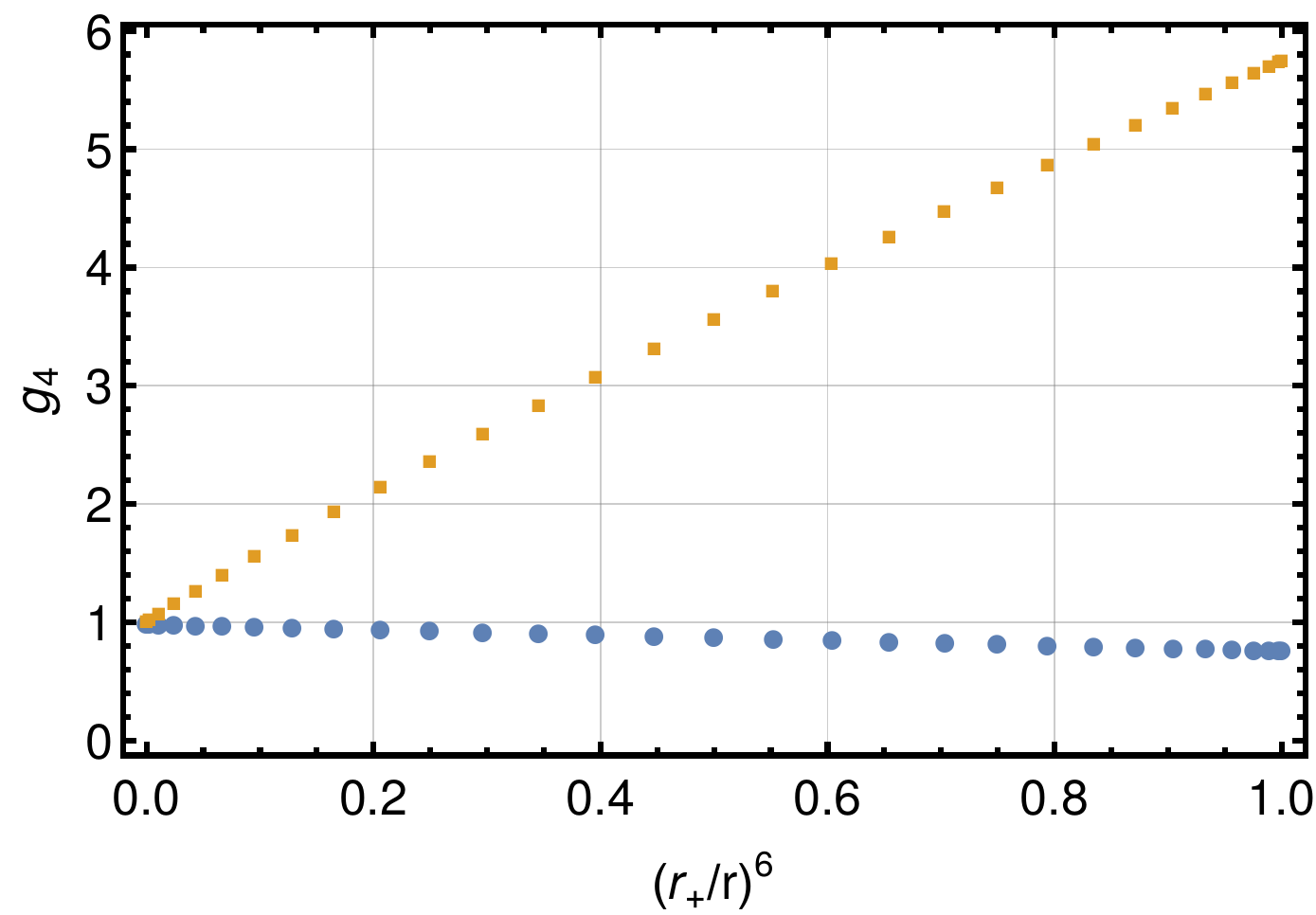}
\end{center}
\caption{The perturbation function $g_4$ for $\beta = 1$ and $\varepsilon = 1$ (lower curve) and $\varepsilon = -1$ (upper curve).
\label{fig:M2influx}}
\end{figure}

As before, we were unable to find analytic solutions for the general finite temperature case, and will therefore turn to numerical methods. To facilitate the numerical evaluation, convert to the compactified coordinate $x = (r_+/r)^6$, for which the equation becomes
\begin{equation}
\partial_x^2 g_4 + \left( \frac{1+2x+x(4-x) \sinh^2\beta}{3x(x-1)\left(1+x\sinh^2\beta\right) } \right) \partial_x g_4 + \left( \frac{3\varepsilon\sinh2\beta  -2+2(2x-3)\sinh^2\beta}{9x(x-1)\left(1+x\sinh^2\beta\right)} \right) g_4 =0. 
\end{equation}
The boundary conditions we desire are that the perturbation approaches the form of the flux background at infinity $g_4(x=0) = 1$, and that the perturbation be regular at the horizon, which requires
\begin{equation}
\left[ \frac{2}{9} \left(1- 3 \varepsilon \tanh\beta \right) g_4 - \partial_x g_4 \right]_{x=1} = 0.
\end{equation}
Once again, \textit{a solution exists for either sign of the charge!} A typical solution is plotted in Fig.~\ref{fig:M2influx}.

The flux is larger at the horizon for $\varepsilon = -1$ than for $\varepsilon = 1$. The norm of the perturbation is
\begin{equation}
|\delta G_4|^2 = 24 m^2 \frac{g_4(r_+)^2}{\cosh^{8/3}\beta}. 
\end{equation}

\begin{figure}[]
\begin{center}
\includegraphics[width=0.5\textwidth]{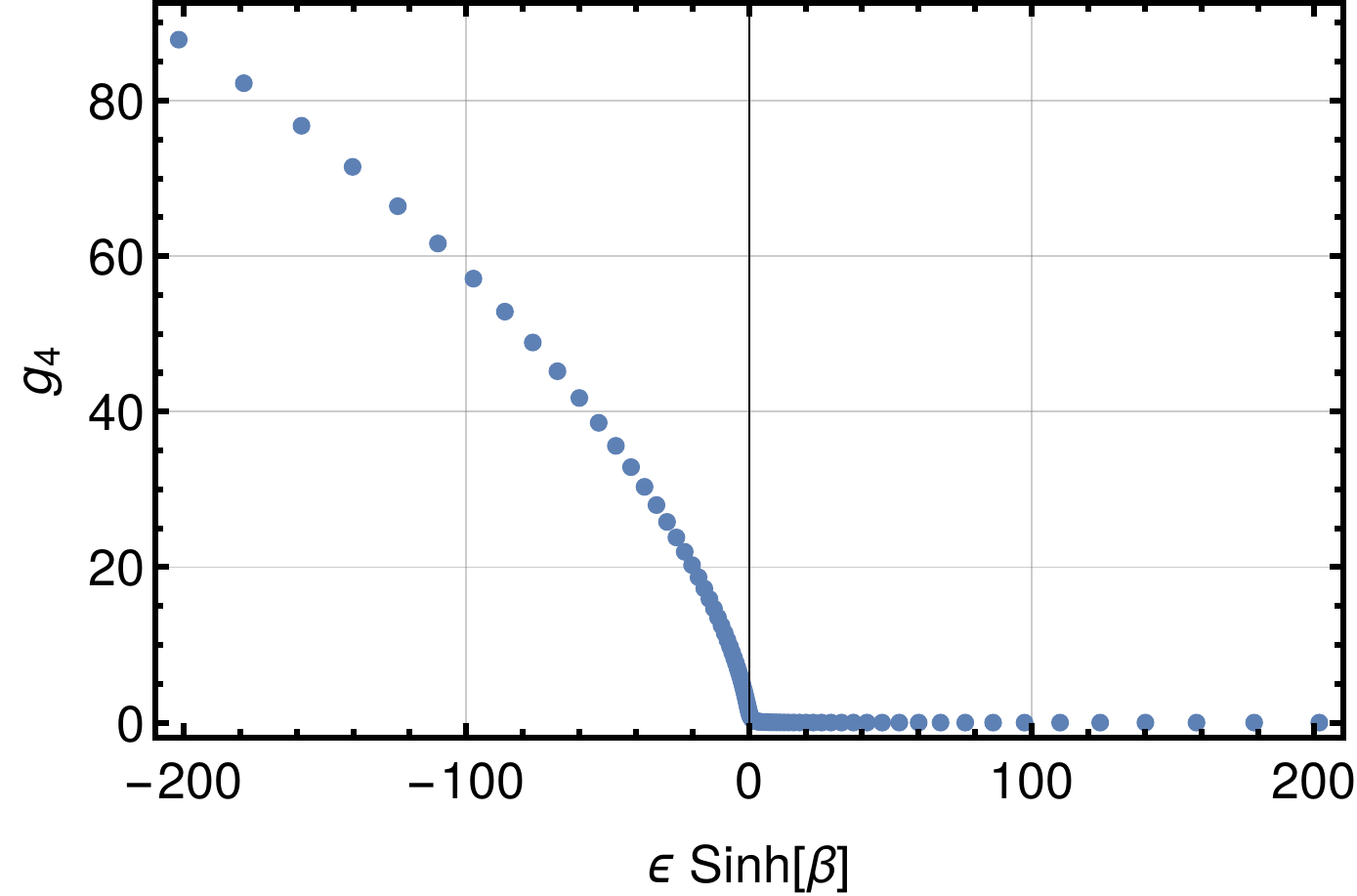}
\end{center}
\caption{The value of the perturbation at the horizon. As extremality is approached, the flux either vanishes $(\varepsilon = 1)$, or diverges $(\varepsilon = -1)$. Recall that our convention is that $\beta\ge 0$ always. Fitting the data indicates that $g_+(r_+) \sim e^{2\beta/3}$ as $\beta \rightarrow \infty$ for $\varepsilon = -1$, which leads to the same growth as in the Type IIB case, $g_4(r_+) \sim e^{2\beta}$.
\label{fig:q1athorizonM2}}
\end{figure}

In Fig.~\ref{fig:q1athorizonM2} the rescaled function $g_+ \equiv g_4(r_+) /\cosh^{4/3}\beta \propto |\delta G_4|$ is plotted against $\beta$ for both signs of $\varepsilon$. As expected, for $\varepsilon \beta \rightarrow + \infty$ the norm vanishes, and as  $\varepsilon \beta \rightarrow - \infty$, the norm diverges.

Once again, an analytic solution exists for the special case of $\beta =0$:
\begin{equation} 
g_4(x) = \, _2F_1\left(-\frac{2}{3},\frac{1}{3};-\frac{1}{3};x\right) - \left( \frac{\Gamma\left(-\frac{1}{6}\right) \Gamma \left( \frac{2}{3}\right)}{ 2^{8/3} \Gamma\left(-\frac{2}{3}\right)\Gamma\left(\frac{7}{6}\right)} \right) x^{4/3} \, _2F_1\left(\frac{2}{3},\frac{5}{3};\frac{7}{3};x\right), 
\end{equation}
and for $\beta = 0$ our numerics agrees with this analytic result. It is also easy to see that the perturbation is consistent with the Raychaudhuri equation, $T_{ab} k^a k^b = 0$. Unlike the Type IIB case, there is no no-go theorem excluding these solutions.

According to the discussion of Sec.~\ref{subsubsec:IIBlocalisedfinitetemp}, the ability to shield the anti-brane singularities behind a finite temperature horizon indicates that the singularities are physical and resolvable by string theory. As before, the natural resolution mechanism is polarisation. To investigate this possibility, we now turn to the extremal case.

\subsubsection{Extremal Case}
For the extremal case, the perturbation equations again have very simple analytic solutions. They are
\begin{align} 
(\varepsilon = 1) & \qquad g_4 = c_0 + c_1 \left(\frac{r_0^8}{r^8} + \frac{4}{7} \frac{r_0^{14}}{r^{14}} \right), \\
(\varepsilon = -1) & \qquad g_4 = c_0 \left(1 + 4 \frac{r_0^6}{r^6} \right) + c_1\frac{r_0^8}{r^8}, \nonumber
\end{align}
where $c_0$, $c_1$ are integration constants. The boundary condition $g_4(\infty) = 1$ requires $c_0 = 1$. Generically, the perturbation is singular in the IR. The norms of the 3-form fluxes are
\[ (\varepsilon = 1) \qquad |G_4|^2 \sim  r^{-20}, \quad c_1 \neq 0, \qquad |G_4|^2 \sim r^8, \quad c_1 = 0. \]
\[ (\varepsilon = -1) \qquad |G_4|^2  \sim r^{-8}, \quad c_1 \neq 0, \qquad |G_4|^2 \sim r^{-4}, \quad c_1 = 0. \]
For $\varepsilon = 1$, the norm can be made finite by setting $c_1 = 0$. As in the Type IIB case, an exact solution exists at the non-linear level corresponding to extremal branes added to the flux background.

For $\varepsilon = -1$, there is no way to keep the asymptotic condition $c_0 = 1$ and also keep the norm finite. The fluxes must diverge as $r \rightarrow 0$. To examine the possibility of the singularity being resolved by polarisation, we first recall that the M2 branes source an $AdS_4$ throat: taking the near-brane extremal limit of the M2 brane solution \eqref{eq:M2} results in the line element
\begin{equation} 
ds^2 \approx \left( \frac{r}{r_0} \right)^4 dx_{\mu}dx^{\mu} + \left( \frac{r_0}{r} \right)^2 dr^2 + r_0^2 d\Omega_7^2.
\end{equation}
After converting to a new radial coordinate $r^2 = 2 r_0 R$, and setting $r_0 = 2 L$, this is seen to be simply $AdS_4 \times S^7$: 
\begin{equation} 
ds^2 \approx \frac{R^2}{L^2} dx_{\mu}dx^{\mu} + \frac{L^2}{R^2}  dR^2 + (2L)^2 d\Omega_7^2.
\end{equation}
The set-up is exactly as in the Type IIB case; the branes source an $AdS$ throat, into which the flux leaks. Depending on the form of the fluxes, polarisation may or may not occur. The M-theory analogue of Polchinski-Strassler was worked out by Bena in \cite{Bena:2000zb}. For $G_4$ perturbations transforming under the transverse $SO(8)$, there are two possible representations, the $\textbf{35}_{\pm}$, whose associated conformal dimensions are:
\begin{align}
\textbf{35}_{+}:& \qquad \Delta = 5, \qquad \Delta = -2, \\ 
\textbf{35}_{-}:& \qquad \Delta = 2, \qquad \Delta = 1, \nonumber 
\end{align}
where again, $\delta G_4 \sim  \lambda^{-\Delta} \delta G_4$ under the scaling of the $AdS_4$ coordinate $R \rightarrow \lambda R$. For $\Delta \le 0$ the mode is regular in the IR $(R \rightarrow 0)$, and for $\Delta > 0$ it diverges. Bena  studied the perturbation corresponding to $\Delta = 1$ in the supersymmetric case and found that the singularity was cured through the physics of brane polarisation.

Returning now to the case of $G_4$ perturbations of the full M2 brane solution, we see that for the case $\varepsilon = -1$, there are three modes present, that is, $\delta G_4$ is a sum of three terms with distinct and definite scalings:
\begin{equation}
\delta G_4 \sim c_1 \lambda^{-2} + c_0 \left( \lambda^{-1} + \lambda^{2} \right).
\end{equation}
The perturbation therefore contains the conformal dimensions $\Delta \in (2,1,-2)$. Taking $c_0 =1$ as required by the boundary condition at infinity forces the $\Delta = 1,-2$ modes to be turned on. The more singular $\Delta = 2$ mode can be discarded through the boundary condition $c_1 =0$. In the deep IR the $\Delta = -2$  mode becomes insignificant and the IR physics is dominated by the $\Delta = 1$. As in the Type IIB case, in order to definitively address whether polarisation occurs, a non-supersymmetric version of \cite{Bena:2000zb} will be needed, which will require the solution to be constructed to higher order. 

To summarise, we find that in both Type IIB string theory and M-theory, regular localised black branes of either sign charge exist for finite temperature. According to the Gubser criterion then, the singularities should be regarded as physical, and are resolved in string theory. In the extremal case of anti-branes added to flux backgrounds the solutions are singular in a manner that is strongly suggests resolution by brane polarisation.

\end{appendix}

\bibliography{refs}{}

\providecommand{\href}[2]{#2}\begingroup\raggedright\begin{thebibliography}{10}

\bibitem{DanielssonXYZ}
U.~Danielsson, S.~Massai, and T.~Van~Riet, ``{SUSY-breaking in warped throats,
  to appear},''.

\bibitem{Kachru:2002gs}
S.~Kachru, J.~Pearson, and H.~L. Verlinde, ``{Brane / flux annihilation and the
  string dual of a nonsupersymmetric field theory},'' {\em JHEP} {\bf 0206}
  (2002) 021, \href{http://xxx.lanl.gov/abs/hep-th/0112197}{{\tt
  hep-th/0112197}}.

\bibitem{Klebanov:2000hb}
I.~R. Klebanov and M.~J. Strassler, ``{Supergravity and a confining gauge
  theory: Duality cascades and chi SB resolution of naked singularities},''
  {\em JHEP} {\bf 0008} (2000) 052,
  \href{http://xxx.lanl.gov/abs/hep-th/0007191}{{\tt hep-th/0007191}}.

\bibitem{Kachru:2003aw}
S.~Kachru, R.~Kallosh, A.~D. Linde, and S.~P. Trivedi, ``{De Sitter vacua in
  string theory},'' {\em Phys.Rev.} {\bf D68} (2003) 046005,
  \href{http://xxx.lanl.gov/abs/hep-th/0301240}{{\tt hep-th/0301240}}.

\bibitem{Argurio:2007qk}
R.~Argurio, M.~Bertolini, S.~Franco, and S.~Kachru, ``{Meta-stable vacua and
  D-branes at the conifold},'' {\em JHEP} {\bf 0706} (2007) 017,
  \href{http://xxx.lanl.gov/abs/hep-th/0703236}{{\tt hep-th/0703236}}.

\bibitem{Argurio:2006ny}
R.~Argurio, M.~Bertolini, S.~Franco, and S.~Kachru, ``{Gauge/gravity duality
  and meta-stable dynamical supersymmetry breaking},'' {\em JHEP} {\bf 0701}
  (2007) 083, \href{http://xxx.lanl.gov/abs/hep-th/0610212}{{\tt
  hep-th/0610212}}.

\bibitem{Bena:2011fc}
I.~Bena, A.~Puhm, and B.~Vercnocke, ``{Metastable Supertubes and non-extremal
  Black Hole Microstates},'' {\em JHEP} {\bf 1204} (2012) 100,
  \href{http://xxx.lanl.gov/abs/1109.5180}{{\tt 1109.5180}}.

\bibitem{Bena:2012zi}
I.~Bena, A.~Puhm, and B.~Vercnocke, ``{Non-extremal Black Hole Microstates:
  Fuzzballs of Fire or Fuzzballs of Fuzz ?},'' {\em JHEP} {\bf 1212} (2012)
  014, \href{http://xxx.lanl.gov/abs/1208.3468}{{\tt 1208.3468}}.

\bibitem{McGuirk:2009xx}
P.~McGuirk, G.~Shiu, and Y.~Sumitomo, ``{Non-supersymmetric infrared
  perturbations to the warped deformed conifold},'' {\em Nucl.Phys.} {\bf B842}
  (2011) 383--413, \href{http://xxx.lanl.gov/abs/0910.4581}{{\tt 0910.4581}}.

\bibitem{Bena:2009xk}
I.~Bena, M.~Grana, and N.~Halmagyi, ``{On the Existence of Meta-stable Vacua in
  Klebanov-Strassler},'' {\em JHEP} {\bf 1009} (2010) 087,
  \href{http://xxx.lanl.gov/abs/0912.3519}{{\tt 0912.3519}}.

\bibitem{Bena:2011hz}
I.~Bena, G.~Giecold, M.~Grana, N.~Halmagyi, and S.~Massai, ``{On Metastable
  Vacua and the Warped Deformed Conifold: Analytic Results},'' {\em
  Class.Quant.Grav.} {\bf 30} (2013) 015003,
  \href{http://xxx.lanl.gov/abs/1102.2403}{{\tt 1102.2403}}.

\bibitem{Massai:2012jn}
S.~Massai, ``{A Comment on anti-brane singularities in warped throats},''
  \href{http://xxx.lanl.gov/abs/1202.3789}{{\tt 1202.3789}}.

\bibitem{Bena:2012bk}
I.~Bena, M.~Grana, S.~Kuperstein, and S.~Massai, ``{Anti-D3 Branes: Singular to
  the bitter end},'' {\em Phys.Rev.} {\bf D87} (2013), no.~10 106010,
  \href{http://xxx.lanl.gov/abs/1206.6369}{{\tt 1206.6369}}.

\bibitem{Blaback:2011nz}
J.~Blaback, U.~H. Danielsson, D.~Junghans, T.~Van~Riet, T.~Wrase, {\em
  et.~al.}, ``{The problematic backreaction of SUSY-breaking branes},'' {\em
  JHEP} {\bf 1108} (2011) 105, \href{http://xxx.lanl.gov/abs/1105.4879}{{\tt
  1105.4879}}.

\bibitem{Blaback:2011pn}
J.~Blaback, U.~H. Danielsson, D.~Junghans, T.~Van~Riet, T.~Wrase, {\em
  et.~al.}, ``{(Anti-)Brane backreaction beyond perturbation theory},'' {\em
  JHEP} {\bf 1202} (2012) 025, \href{http://xxx.lanl.gov/abs/1111.2605}{{\tt
  1111.2605}}.

\bibitem{Blaback:2014tfa}
J.~Blaback, U.~Danielsson, D.~Junghans, T.~Van~Riet, and S.~Vargas,
  ``{Localised anti-branes in non-compact throats at zero and finite T},''
  \href{http://xxx.lanl.gov/abs/1409.0534}{{\tt 1409.0534}}.

\bibitem{Myers:1999ps}
R.~C. Myers, ``{Dielectric branes},'' {\em JHEP} {\bf 9912} (1999) 022,
  \href{http://xxx.lanl.gov/abs/hep-th/9910053}{{\tt hep-th/9910053}}.

\bibitem{Polchinski:2000uf}
J.~Polchinski and M.~J. Strassler, ``{The String dual of a confining
  four-dimensional gauge theory},''
  \href{http://xxx.lanl.gov/abs/hep-th/0003136}{{\tt hep-th/0003136}}.

\bibitem{Klebanov:2000nc}
I.~R. Klebanov and A.~A. Tseytlin, ``{Gravity duals of supersymmetric SU(N) x
  SU(N+M) gauge theories},'' {\em Nucl.Phys.} {\bf B578} (2000) 123--138,
  \href{http://xxx.lanl.gov/abs/hep-th/0002159}{{\tt hep-th/0002159}}.

\bibitem{Horowitz:1995ta}
G.~T. Horowitz and R.~C. Myers, ``{The value of singularities},'' {\em
  Gen.Rel.Grav.} {\bf 27} (1995) 915--919,
  \href{http://xxx.lanl.gov/abs/gr-qc/9503062}{{\tt gr-qc/9503062}}.

\bibitem{Bena:2012tx}
I.~Bena, D.~Junghans, S.~Kuperstein, T.~Van~Riet, T.~Wrase, {\em et.~al.},
  ``{Persistent anti-brane singularities},'' {\em JHEP} {\bf 1210} (2012) 078,
  \href{http://xxx.lanl.gov/abs/1205.1798}{{\tt 1205.1798}}.

\bibitem{Bena:2012vz}
I.~Bena, M.~Grana, S.~Kuperstein, and S.~Massai, ``{Polchinski-Strassler does
  not uplift Klebanov-Strassler},'' {\em JHEP} {\bf 1309} (2013) 142,
  \href{http://xxx.lanl.gov/abs/1212.4828}{{\tt 1212.4828}}.

\bibitem{Bena:2014bxa}
I.~Bena, M.~Grana, S.~Kuperstein, and S.~Massai, ``{Tachyonic Anti-M2
  Branes},'' {\em JHEP} {\bf 1406} (2014) 173,
  \href{http://xxx.lanl.gov/abs/1402.2294}{{\tt 1402.2294}}.

\bibitem{Bena:2014jaa}
I.~Bena, M.~Grana, S.~Kuperstein, and S.~Massai, ``{Giant Tachyons in the
  Landscape},'' \href{http://xxx.lanl.gov/abs/1410.7776}{{\tt 1410.7776}}.

\bibitem{Junghans:2014wda}
D.~Junghans, D.~Schmidt, and M.~Zagermann, ``{Curvature-induced Resolution of
  Anti-brane Singularities},'' {\em JHEP} {\bf 1410} (2014) 34,
  \href{http://xxx.lanl.gov/abs/1402.6040}{{\tt 1402.6040}}.

\bibitem{Gubser:2000nd}
S.~S. Gubser, ``{Curvature singularities: The Good, the bad, and the naked},''
  {\em Adv.Theor.Math.Phys.} {\bf 4} (2000) 679--745,
  \href{http://xxx.lanl.gov/abs/hep-th/0002160}{{\tt hep-th/0002160}}.

\bibitem{Bena:2012ek}
I.~Bena, A.~Buchel, and O.~J. Dias, ``{Horizons cannot save the Landscape},''
  {\em Phys.Rev.} {\bf D87} (2013), no.~6 063012,
  \href{http://xxx.lanl.gov/abs/1212.5162}{{\tt 1212.5162}}.

\bibitem{Bena:2013hr}
I.~Bena, J.~Blaback, U.~Danielsson, and T.~Van~Riet, ``{Antibranes cannot
  become black},'' {\em Phys.Rev.} {\bf D87} (2013), no.~10 104023,
  \href{http://xxx.lanl.gov/abs/1301.7071}{{\tt 1301.7071}}.

\bibitem{Blaback:2012nf}
J.~Blaback, U.~H. Danielsson, and T.~Van~Riet, ``{Resolving anti-brane
  singularities through time-dependence},'' {\em JHEP} {\bf 1302} (2013) 061,
  \href{http://xxx.lanl.gov/abs/1202.1132}{{\tt 1202.1132}}.

\bibitem{Danielsson:2014yga}
U.~H. Danielsson and T.~Van~Riet, ``{Fatal attraction: more on decaying
  anti-branes},'' \href{http://xxx.lanl.gov/abs/1410.8476}{{\tt 1410.8476}}.

\bibitem{Cvetic:2000db}
M.~Cvetic, G.~Gibbons, H.~Lu, and C.~Pope, ``{Ricci flat metrics, harmonic
  forms and brane resolutions},'' {\em Commun.Math.Phys.} {\bf 232} (2003)
  457--500, \href{http://xxx.lanl.gov/abs/hep-th/0012011}{{\tt
  hep-th/0012011}}.

\bibitem{Michel:2014lva}
B.~Michel, E.~Mintun, J.~Polchinski, A.~Puhm, and P.~Saad, ``{Remarks on brane
  and antibrane dynamics},'' \href{http://xxx.lanl.gov/abs/1412.5702}{{\tt
  1412.5702}}.

\bibitem{Janssen:1999sa}
B.~Janssen, P.~Meessen, and T.~Ortin, ``{The D8-brane tied up: String and brane
  solutions in massive type IIA supergravity},'' {\em Phys.Lett.} {\bf B453}
  (1999) 229--236, \href{http://xxx.lanl.gov/abs/hep-th/9901078}{{\tt
  hep-th/9901078}}.

\bibitem{Horowitz:2000kx}
G.~T. Horowitz and V.~E. Hubeny, ``{Note on small black holes in AdS(p) x
  S**q},'' {\em JHEP} {\bf 0006} (2000) 031,
  \href{http://xxx.lanl.gov/abs/hep-th/0005288}{{\tt hep-th/0005288}}.

\bibitem{Freedman:2000xb}
D.~Z. Freedman and J.~A. Minahan, ``{Finite temperature effects in the
  supergravity dual of the N=1* gauge theory},'' {\em JHEP} {\bf 0101} (2001)
  036, \href{http://xxx.lanl.gov/abs/hep-th/0007250}{{\tt hep-th/0007250}}.

\bibitem{Zamora:2000ha}
F.~Zamora, ``{Nonsupersymmetric SO(3) invariant deformations of N=1* vacua and
  their dual string theory description},'' {\em JHEP} {\bf 0012} (2000) 021,
  \href{http://xxx.lanl.gov/abs/hep-th/0007082}{{\tt hep-th/0007082}}.

\bibitem{DeWolfe:2004qx}
O.~DeWolfe, S.~Kachru, and H.~L. Verlinde, ``{The Giant inflaton},'' {\em JHEP}
  {\bf 0405} (2004) 017, \href{http://xxx.lanl.gov/abs/hep-th/0403123}{{\tt
  hep-th/0403123}}.

\bibitem{Dymarsky:2011pm}
A.~Dymarsky, ``{On gravity dual of a metastable vacuum in Klebanov-Strassler
  theory},'' {\em JHEP} {\bf 1105} (2011) 053,
  \href{http://xxx.lanl.gov/abs/1102.1734}{{\tt 1102.1734}}.

\bibitem{Bena:2000zb}
I.~Bena, ``{The M theory dual of a three-dimensional theory with reduced
  supersymmetry},'' {\em Phys.Rev.} {\bf D62} (2000) 126006,
  \href{http://xxx.lanl.gov/abs/hep-th/0004142}{{\tt hep-th/0004142}}.

\bibitem{Klebanov:2010qs}
I.~R. Klebanov and S.~S. Pufu, ``{M-Branes and Metastable States},'' {\em JHEP}
  {\bf 1108} (2011) 035, \href{http://xxx.lanl.gov/abs/1006.3587}{{\tt
  1006.3587}}.

\bibitem{Horowitz:2011cq}
G.~T. Horowitz and T.~Wiseman, ``{General black holes in Kaluza-Klein
  theory},'' \href{http://xxx.lanl.gov/abs/1107.5563}{{\tt 1107.5563}}.

\bibitem{Hartnett:2014zca}
G.~S. Hartnett, G.~T. Horowitz, and K.~Maeda, ``{A No Black Hole Theorem},''
  \href{http://xxx.lanl.gov/abs/1410.1875}{{\tt 1410.1875}}.

\bibitem{Cvetic:2002hi}
M.~Cvetic, H.~Lu, and C.~Pope, ``{Penrose limits, PP waves and deformed M2
  branes},'' {\em Phys.Rev.} {\bf D69} (2004) 046003,
  \href{http://xxx.lanl.gov/abs/hep-th/0203082}{{\tt hep-th/0203082}}.

\bibitem{Horowitz:1990sr}
G.~T. Horowitz and A.~R. Steif, ``{Strings in Strong Gravitational Fields},''
  {\em Phys.Rev.} {\bf D42} (1990) 1950--1959.

\end{thebibliography}\endgroup
\bibliographystyle{utphys}

\end{document}